\documentclass[fleqn,usenatbib]{mnras}

\usepackage[T1]{fontenc}

\DeclareRobustCommand{\VAN}[3]{#2}
\let\VANthebibliography\thebibliography
\def\thebibliography{\DeclareRobustCommand{\VAN}[3]{##3}\VANthebibliography}

\usepackage{graphicx}	
\usepackage{amsmath}	
\usepackage{amssymb}

\usepackage{newtxtext,newtxmath}
\usepackage{lipsum}
\usepackage{xspace}
\usepackage[dvipsnames]{xcolor}

\newcommand{\pipe}{\textsc{Pipe3D}\xspace}
\newcommand{\angstrom}{\text{\normalfont\AA}}
\newcommand{\Msun}{\ensuremath{{\rm M_\odot}}\xspace}
\newcommand{\ha}{\ensuremath{\rm H\alpha}\xspace}
\newcommand{\ew}{\ensuremath{\rm EW(\ha)}\xspace}
\newcommand{\reff}[1]{\ensuremath{#1\,R_{\rm e}}\xspace}
\newcommand{\dd}{{\rm d}}

\newcommand{\balbreak}{\ensuremath{D_n(4000)}\xspace}
\newcommand{\gr}{\ensuremath{(g-r)_0}\xspace}
\newcommand{\tshort}{\ensuremath{20}\xspace}
\newcommand{\tlong}{\ensuremath{3000}\xspace}
\newcommand{\massshort}{\ensuremath{M_*^{\tshort}/M_*^{\rm tot}}\xspace}
\newcommand{\masslong}{\ensuremath{M_*^{\tlong}/M_*^{\rm tot}}\xspace}
\newcommand{\taudeath}{\ensuremath{\tau_{\rm death}}\xspace}

\newcommand{\change}[1]{#1}
\newcommand{\cchange}[1]{#1}
\newcommand{\ccchange}[1]{#1}

\defcitealias{Corcho-Caballero+20}{CC20}
\defcitealias{Corcho-Caballero+21b}{CC21b}
\defcitealias{Casado+15}{C15}

\title[Ageing and Quenching through the ageing diagram]{Ageing and Quenching through the ageing diagram: predictions from simulations and observational constraints.}

\author[P. Corcho-Caballero et al.]{
	Pablo Corcho-Caballero,$^{1,2,3}$\thanks{E-mail: pablo.corcho@uam.es}
	Yago Ascasibar,$^{1}$
	Sebastián F. S\'anchez,$^{4}$
    and
	\'Angel R. L\'opez-S\'anchez$^{2,3,5}$
	\\
	$^{1}$Departamento de Física Teórica, Universidad Autónoma de Madrid (UAM), Campus de Cantoblanco, Madrid 28049, Spain\\
	$^{2}$Australian Astronomical Optics, Macquarie University, 105 Delhi Rd, North Ryde, NSW 2113, Australia\\
	$^{3}$ARC Centre of Excellence for All Sky Astrophysics in 3 Dimensions (ASTRO-3D)\\
	$^{4}$Instituto de astronom\'ia, Universidad Nacional Aut\'onoma de M\'exico, A.P. 70-264, 04510 M\'exico D. F., M\'exico\\
	$^{5}$Macquarie University Research Centre for Astronomy, Astrophysics \& Astrophotonics, Sydney, NSW 2109, Australia.
}

\date{Accepted 2023 January 11. Received 2023 January 8; in original form 2022 August 26}

\pubyear{2022}

\begin{document}
\label{firstpage}
\pagerange{\pageref{firstpage}--\pageref{lastpage}}
\maketitle

\begin{abstract}
\change{We study recent changes on the star formation history (SFH) of galaxies by means of the ageing diagram (AD), tracing the fraction of stars formed during the last \cchange{$\sim \tshort$~Myr} through the equivalent width of the \ha line and $\sim 1-3$~Gyr through the dust-corrected optical colour \gr or the Balmer break.
We provide a physical characterization by using \pipe estimates of the SFH of CALIFA and MaNGA galaxies, in combination with the predictions from IllustrisTNG-100.
Our results show that the AD may be divided into four domains that correlate with the stellar mass fractions formed in the last \cchange{\tshort Myr} and 3 Gyr:}
Ageing systems, whose SFR changes on scales \change{of several Gyr},
account for $70-80\%$ of the galaxy population. 
Objects whose SFH was abruptly truncated \change{in the last $\sim 1$ Gyr} arrange along a detached Quenched sequence that represents $\sim 5-10\%$ by (volume-corrected) number for $10^9 < M_*/\Msun < 10^{12}$.
Undetermined systems represent an intermediate population between the Ageing and Quenched regimes.
Finally, Retired galaxies, \change{dominated by old stellar populations}, are located at the region in the AD where the Ageing and Quenched sequences converge.
Defining different star formation activity levels in terms of the birth rate parameter $b\equiv \frac{SFR}{\langle SFR \rangle}$, we find that galaxies transit from the Ageing to Quenched sequences on scales $\sim 500$~Myr.
We conclude that the ageing diagram provides a useful tool to discern recently Quenched galaxies from the dominant Ageing population.
\end{abstract}




\section{Introduction}
\label{sec:intro}
Galaxies often exhibit bimodal distributions in terms of observable properties like color \citep{Strateva+01, Blanton+03, Baldry+04, Baldry+06, Taylor+15}, morphology \citep{Bamford+09} or physical quantities such as stellar kinematics \citep{Emsellem+11, Cappellari+16, Smethurst+18} and star formation rate \citep{Brinchmann+04, Peng+10, Moustakas+13}.
This has led to the broad division of galaxies into two distinct populations: \emph{blue cloud} (BC) or \emph{star-forming} galaxies, presenting disk morphologies with abundant gas reservoirs and \emph{red sequence} (RS) or \emph{passive}\footnote{Also referred to as retired or quenched.} galaxies that typically correspond to gas-poor early-type systems.

The transition from the BC to the RS and thus the demise of star formation in galaxies is often attributed to some \emph{quenching} process, able to truncate the conversion of gas into stars in short timescales with respect to the age of the Universe \citep[e.g.][]{Schawinski+14}.
The most popular scenarios for quenching star-forming satellite galaxies are related to environmental (hydrodynamically triggered) processes such as ram pressure stripping in dense clusters \citep[e.g.][]{Gunn&Gott72, Boselli&Gavazzi06, Brown+17}, strangulation/starvation \citep[e.g.][]{Wetzel+13, Peng+15} or (gravitationally triggered) galaxy harassment \citep[e.g.][]{Moore+96, Bialas+15}, see \citet{Cortese+21} for a recent review on environmental \mbox{quenching.}
On the other hand, `quenching' in central galaxies is thought to be driven by a wide variety of internal processes, such as negative feedback produced either by star formation \citep[e.g.][]{MacLow&Ferrara99, Stinson+09, Sawala+10, Fitts+17} or active galactic nuclei (AGN) \citep{Croton+06, Cheung+16, Martin-Navarro+21}, as well as gas stabilization against cloud fragmentation \citep{Bigiel+08, Martig+09, Ceverino+10}.
It is important to note that the timescales involved range between a few hundreds of Myr to several Gyr \citep{Peng+10, Wetzel+13, Hirschmann+14, Hahn+17, Wright+19, Walters+22}, although the precise values are subjected to the adopted definition of the timescale.

Whether there are two channels (fast and slow) for truncating the star-formation history (SFH) of galaxies is currently under debate.
Typically, the amount of time required for truncating the star formation history of a galaxy has been characterized by means of the time spent crossing two arbitrary boundaries over a given parameter space -- e.g. colour-colour or colour-magnitude diagram \citep{Phillips+19, Wright+19, Akins+22} and the stellar mass ($M_*$) -- specific star-formation rate ($sSFR = SFR/{M_*}$) plane \citep{Wetzel+13, Hahn+17, Tacchella+22}.
In general, those processes terminating the SFH of galaxies in less than $\lesssim 1$ Gyr are regarded as `fast quenching', while the term `slow quenching' is used to denote processes with timescales comparable to the age of the Universe \citep[e.g.][]{Tacchella+22, Suess+22b}.

In our opinion, the word `quenching' should be reserved to \emph{discrete} events that act upon a star-forming galaxy and make it migrate to the RS on a certain timescale (hence `fast' or `slow' quenching).
In contrast, we use the term `ageing' to denote smooth secular evolution throughout the life of a galaxy, without any singular event (very different, but difficult to distinguish from slow quenching in practice).

Moreover, some studies even question the existence of two different physical states.
In the bimodal paradigm, there is a `main sequence' (MS) of star-forming galaxies \citep{Noeske+07} that describes the location of these systems in the $M_*-sSFR$ plane (analogous to the blue cloud), completely detached from the population of passive galaxies whose specific star-formation rate is negligible, if not zero.
Actually, the separation in two discrete states is rather arbitrary unless passive galaxies are truly `dead' \citep{Corcho-Caballero+21a}.
The sole cold gas exhaustion, driven only by the conversion of gas into stars, following a continuous and decreasing star formation history, is able to recover the main features observed in the $M_*-sSFR$ relation \citep{Oemler+17}.
\citet{Eales+18a, Eales+18b} showed that the majority of red star-forming galaxies in optical surveys are classified as passive systems.
This interpretation was reinforced in \citet{Corcho-Caballero+20} (hereafter CC20), were a thorough statistical analysis in terms of $M_*$ and $sSFR$ provided a picture of galaxies arranging along a (intrinsically bidimensional) distribution with a single mode corresponding to the MS and extended power-law tails accounting for the `passive' and `starburst' populations.

However, recovering the star formation histories of RS galaxies from optical data is extremely challenging and requires very high signal-to-noise ratios and/or multiwavelength approaches that usually hinder a statistical analysis in terms of sample completeness \citep{Sanchez+19a, Salvador-Rusinol+20, deLorenzo-Caceres+20, Camps-Farina+22}.
In addition, theoretical predictions using numerical simulations are still sensitive to resolution and subgrid physics \citep[e.g.][]{Schaye+15, Zhao+20} that are key to distinguish between both scenarios.
In fact, some recent studies provided evidence of an apparent discrepancy on the passive regime between observational estimates and state-of-the-art cosmological simulations \citep{Katsianis+21, Corcho-Caballero+21a}, suggesting that the feedback implementation might be preventing galaxies from retaining low levels of star formation.

Following the our adopted definition of quenching, the most likely candidates of quenched systems would correspond to the so-called post-starburst (PSB) galaxies (PSBGs) \citep[e.g.][]{Dressler&Gunn83, Poggianti+99, Balogh+05}, also referred to as E+A or K+A; i.e. systems with prominent Balmer absorption lines --$\rm EW(H\delta) < 5$ \AA-- but mild or null nebular emission (typically traced by the \ha and/or $\rm [\ion{O}{ii}]\lambda3727~\angstrom$ lines).
This systems are thought to represent a transition population that underwent some quenching episode recently (sometimes following a burst of star-formation), with the consequent demise of O and B stars in favour of the A/F type, giving rise to their characteristic optical spectrum.

During the last few decades, a lot of effort has been put into determining what are the properties of this transitory phase and what are the main drivers that quench PSBGs \citep[e.g.][and references therein]{Goto+07, DiMatteo+08, Dressler+13, Pawlik+18, Pawlik+19}.
Recently, \citet{Chen+19} analyzed $\sim300$ PSBGs drawn from a sample of $\sim 4000$ MaNGA galaxies, and found a positive radial gradient in terms of optical features such as the Balmer Break or EW(H$\delta$); suggesting a strong gas fuelling towards the central regions that induce severe star formation bursts, followed by a halt of the SFH due to gas exhaustion.
In the same line, other studies \citep[e.g.][]{Owers+19, Zheng+20, Zheng+22} support the scenario of field PSBGs being quenched through mergers followed by black-hole (BH) feedback, while quenching in dense environments might be driven by hydrodynamical processes like ram pressure stripping.
The fraction of PSGBs is indeed extremely scarce; accounting only for a few percent in most of the samples \citep[e.g.][]{Goto+07, Alatalo+16}, due to the short-lived period in which galaxies present observational features below the classic thresholds: $\rm EW(H\delta) < 5~\angstrom$ and $\log_{10}(\rm |W(H\alpha)|/\angstrom) < 0.3$\footnote{Throughout this work we will refer to the raw equivalent width, including emission and absorption features, to as EW; while equivalent widths derived using only emission lines will be regarded as W.}.

Complementary to the previous approach, the \emph{ageing diagram} (AD) \citep[][C15 and CC21b, respectively]{Casado+15, Corcho-Caballero+21b} provides a powerful tool to study quenching processes on galaxies.
By means of two observational proxies of star formation, sensitive to different timescales, it is possible to trace recent changes on the star-formation history of galaxies.
In particular, the AD vertical axis represents an estimate of the average $sSFR$ over the last few Myr, such as the \ew.
This quantity traces the mass fraction of young stellar populations, whose emission is dominated by the presence of massive O and B stars, that are able to ionise the interstellar medium (ISM).
Additionally, the horizontal axis reflects the average specific star formation rate over longer timescales ($\sim0.1-1$ Gyr), usually traced by broad-band colors, e.g. $(u-r)$ in \citetalias{Casado+15} and $(g-r)$ in \citetalias{Corcho-Caballero+21b} or the Balmer Break at 4000\AA, \balbreak, in this work; that correspond to the fraction of intermediate age stellar populations, dominated by A-type stars.

As a result, this diagram is able to discriminate between two evolutionary pathways.
On the one hand, those systems dominated by secular evolution, with slowly decreasing star-formation histories, will present small variations on the SFH during the last few hundred Myr.
They will arrange across the \emph{ageing sequence}, presenting different levels of SF activity: from high \ew and blue continuum features (BC) to low level of \ha emission and red colours (RS).
In contrast, galaxies that suffered quenching during the last $\sim$ Gyr will appear as a detached population; lining up along the \emph{quenched sequence}, from strong \ha absorption and blue colours towards gradually decreasing absorption features and redder spectra, according to the time at which quenching occurred.

Building upon previous studies \citepalias{Casado+15, Corcho-Caballero+21b}, in this present work we aim to provide a comprehensive physical characterisation of ageing diagram in terms of the star formation history of galaxies.
The paper will be divided as follows: Section~\ref{sec:obs_data}  describes the observational galaxy samples used in the present work and the adopted empirical star-formation proxies.
Section~\ref{sec:models} describes the theoretical star-formation histories in Sect.~\ref{sec:pipe3d_tng_sfhs} and the model to predict the \ha emission. 
\cchange{Our results are presented in Sect.~\ref{sec:results}, the AD characterization is described in Sect.~\ref{sec:the_ageing_diagram}, the results for IllustrisTNG are shown described in Sect.~\ref{sec:AD_classification} and the statistical results for the observational sample are presented in Sect.~\ref{sec:AD_statistics}.}
Section~\ref{sec:discussion} discusses the connection of the distribution of galaxies along the ageing diagram with physical properties: SFH (Sect.~\ref{sec:star_formation_histories}), evolutionary timescales (Sect.~\ref{sec:timescales}) and young stellar mass fractions (Sect.~\ref{sec:mass_formed}).
Finally, we provide a summary and our main conclusions on Sect.~\ref{sec:conclusions}.

Throughout this work, we adopt a $\Lambda$CMD cosmology, with $\rm H_0=70~ km/s/Mpc$ and $\Omega_m = 0.3$.

\section{Observational Data}
\label{sec:obs_data}

\subsection{Galaxy samples}
Studying the current evolutionary status of galaxies is challenging in multiple ways.
The fundamental issue resides on the need for accurate measurements with high SNR for inferring the fundamental properties of the multiple components of the galaxy (cold/warm/hot gas, stellar populations, dust).
Yet, classifying galaxies in terms of a unique global status is also controversial as they appear to be undergoing different evolutionary stages across their surface \citepalias[e.g.][]{Corcho-Caballero+21b}.
Analysis based on single-fiber spectra tend to provide insights of the inner regions of nearby galaxies; where they typically present signatures of being on the late evolutionary stages, but they cannot be considered as representative of the global status of the system.
Therefore, in order to study the global properties of galaxies, it is key to account for a significant spatial extent of their surface in order to include the different stages the system is going through.

Since the advent of Integral Field Spectroscopy (IFS), spatially resolved analysis of nearby galaxies have provided a deeper and new view on galaxy evolution. 
This method allows to obtain from tenths to thousands of spectra per galaxy across their optical extent.
In this work we will select two galaxy samples from two of the largest IFS surveys to date -- CALIFA (\S\ref{sec:CALIFA}) and MaNGA (\S\ref{sec:MANGA}) --
and analyse the optical properties contained within $\rm1.5\,R_{e}$ in stellar mass (see below for details); considering it as representative of the total emitted spectrum of each system.

\subsubsection{CALIFA}
\label{sec:CALIFA}
The third public data release (DR3) of the Calar Alto Legacy Integral Field Area (CALIFA) survey \citep{CALIFApresentation, CALIFADR1, CALIFADR2, CALIFADR3} consist of a sample of nearby diameter-selected galaxies with an excellent optical coverage ($\gtrsim$\reff{2.5}), that span a wide range of kinematic and morphological properties \citep[e.g.][]{Gonzalez-Delgado+15, Falcon-Barroso+17}.
The survey was conducted using the Potsdam Multi-Aperture Spectrophotometer \citep[PMAS,][]{Roth+05} mounted on the 3.5 telescope of the Calar Alto observatory, employing the PPak wide-field integral field unit \citep{Kelz+06} with a field of view $\sim 1~\rm arcmin^2$ composed of 382 fibres of 2.7 arcsecs diameter each.

The mother sample selection criteria is based on a size cut using the isophote major axis at 25 magnitudes per square arsecond in the r band; $45''<isoA_{\rm r}<79.2''$, a redshift limiting range; $0.005 < z < 0.03$ and a flux limit in the $r$ band petrosian magnitude; $r_{\rm petro}< 20$ \citep[see][for details]{CALIFAselection}.
This provides a 95\% completeness limit in the stellar mass range $10^{9.7}-10^{11.4}$ that allows to compute $V_{\rm max}$ values according to \citet{Schmidt68}.
In particular, in this work we will use the low-resolution setup (V500), covering the wavelength range 3745–7500 \AA\, with a spectral resolution of 6 \AA\,(FWHM).
After discarding galaxies with unreliable star formation histories (see details on \S\ref{sec:pipe3d}), the CALIFA sample used in this work comprises 472 objects.

\subsubsection{MaNGA}
\label{sec:MANGA}
The MaNGA (Mapping Nearby Galaxies at Apache Point Observatory) survey \citep{MANGAoverview} is one of the fourth-generation Sloan Digital Sky Survey core programs, which was able to measure spatially resolved spectroscopy of $\sim10000$ galaxies.
The instrument for carrying out the survey employs 17 fiber-bundle integral field units that vary in diameter from $12''$ to $32''$ (19 to 127 fibers per IFU) with a wavelength coverage over $3600-10300$~\AA\, at $R\sim2000$ \citep{Drory+15}.

The first principle that motivated the sample selection consists of getting a large enough number of galaxies to fill six bins of stellar mass, SFR and environment (216 bins) with 50 objects respectively \citep{MANGAselection, Yan+16a}.
Additionally, galaxies are selected following a flat distribution in terms of stellar mass in the range $9 < \log_{10}(M_*/M_\odot)< 12$ (based on the K-corrected $i$-band).
In terms of the spatial extent, approximately two thirds of the total sample were observed up to \reff{1.5} (Primary sample) while $\sim$ one third is covered up to \reff{2.5} (Secondary sample).
In addition, a third sample was selected to overpopulate the number of green valley systems; currently between the star-forming and passive populations, comprising $\sim10\%$ of the total sample.
This considerations translate into a sample of galaxies that ranges in redshift between $0.01 \lesssim z \lesssim 0.15$.

In this work we are using the complete MaNGA release, as part of the 17th SDSS data release \citep{SDSSDR17}, that comprises more than 10.000 galaxies.
After selecting those with reliable measurements of the star formation history (see \S\ref{sec:pipe3d}), we end up with a total sample of 8590 galaxies.
We will use the publicly available weights provided in \citet{Sanchez+22} in order to perform the statistical analysis using a volume-corrected sample.

\subsection{Star formation proxies}

Based on the rest-frame integrated spectrum within \reff{1.5} of total stellar mass (see \ref{sec:models} for details), $F_\lambda$, we will compute the following proxies for star formation:

\begin{itemize}
	\item {Dust-free \gr color index}: computed by convolving each spectra, applying dust extinction corrections in the case of observational spectra, with the corresponding SDSS filter response function $T_\nu$, using the AB system:
	\begin{equation}
	m = -2.5 \log_{10}\left(\frac{\int_0^\infty F_\nu T_\nu \dd\log_{10}(\nu)}{\int_0^\infty T_\nu \dd\log_{10}(\nu)}\right) - 48.60
	\end{equation}
	Where $F_\nu$ denotes the integrated specific flux per frequency unit.
	\item {Balmer break, $D_{n}(4000)$}: computed as the ratio between the average flux at $4050-4250~$\AA\, and $3850-3950~$\AA, respectively:
	\begin{equation}
	D_n(4000) = \frac{\langle F_\lambda \rangle_{4050-4250~\AA}}{\langle F_\lambda \rangle_{3850-3950~\AA}}
	\end{equation}
	\item {Raw \ha equivalent width}, \ew, defined as:
	\begin{equation}
	\label{eq:ew}
	{\rm EW(H\alpha)}
	\equiv
	\int_{\rm 6550\,\AA}^{\rm 6575\,\AA}
	\frac{ F_\lambda(\lambda) }
	{\frac{F_{\rm B}\lambda_{\rm R}-F_{\rm R}\lambda_{\rm B}}{\lambda_{\rm R}-\lambda_{\rm B}}+
		\lambda\frac{F_{\rm R}-F_{\rm B}}{\lambda_{\rm R}-\lambda_{\rm B}}}-1\ \dd\lambda
	\end{equation}
	where $F_{\rm B}$ and $F_{\rm R}$ correspond to the mean flux per unit wavelength computed in  the $6470-6530$~\AA\ and $6600-6660$~\AA\ bands, with central wavelengths $\lambda_{\rm B}=6500$~\AA\ and $\lambda_{\rm R}=6630$~\AA, respectively. 
	Under this definition, positive and negative values of EW denote emission and absorption, respectively.
\end{itemize}

As already mentioned in the introduction, these proxies are sensitive to different timescales, where the first two trace the fraction of stars formed during the last 300-1000 Myr, while \ew probes younger stellar populations, of the order of $\sim1-100$ Myr (see \S\ref{sec:ha_models} for more details).

\section{Models}
\label{sec:models}
As discussed in \citetalias{Casado+15} and \citetalias{Corcho-Caballero+21b} the AD is able to discriminate between smooth and sharply truncated star formation histories.
In this work we will explore the power of this diagram by means of several sets of synthetic observations built upon the star formation histories provided by empirical SED-fitting methods (\S\ref{sec:pipe3d}) and theoretical models (\S\ref{sec:IllustrisTNG}).
We will also address the possible systematic effects that may arise when the \ha\,emission is not only produced by young massive stars (\S\ref{sec:ha_models}) but also from old stellar populations or processes typically associated with these environments.

\subsection{Star formation histories}
\label{sec:pipe3d_tng_sfhs}

\subsubsection{Pipe3D}
\label{sec:pipe3d}
The \pipe pipeline accounts for the stellar continuum of galaxy spectra by fitting a range of simple stellar population (SSP) models, including the features produced by stellar kinematics and dust extinction, as well as the nebular emission analysis \citep{Sanchez+16a, Sanchez+16b, Lacerda+22a}.
The fitting sequence starts by using a small number of SSP templates to determine the non-linear properties, i.e. stellar kinematics and dust extinction \citep[adopting the attenuation curve from][]{Cardelli+89} present in the spectra.
Then, the emission and kinematic properties of the ionized gas are estimated by means of Gaussian fits to each pre-selected line.
Finally, the whole set of stellar templates is used to fit each gas-free spectrum, providing a set of weights from which the star formation history can be recovered. 

In this work we make use of the results obtained using the \texttt{GSD156} SSP library \citep{Cid-Fernandes+13}, applied to both CALIFA and MaNGA samples, that adopts the \citet{Salpeter55} Initial Mass Function (IMF) for stellar masses between 0.1 and 100 $\rm M_\odot$.
This library comprises 156 individual templates, sampling 39 (quasi)log-scaled ages from 1 Myr to 14 Gyr and 4 individual metallicities ($Z/Z_\odot=0.2, 0.4, 1, 1.5$).

When computing the integrated star formation history of each galaxy, we restrict it to a circular aperture of \reff{1.5} based on the projected stellar mass map provided by \pipe, masking the contribution from foreground stars if required.

\subsubsection{IllustrisTNG}
\label{sec:IllustrisTNG}
We use the star formation histories provided by the IllustrisTNG project\footnote{\url{https://www.tng-project.org/}} \citep{Naiman+18, Marinacci+18,Springel+18, Pillepich+18b, Nelson+18}.
This suite comprises a series of cosmological magneto-hidrodynamical simulations, run with the moving-mesh \texttt{AREPO} code \citep{Springel10}, that model a vast range of physical processes such as gas cooling and heating, star formation, stellar evolution and chemical enrichment, SN feedback, BH growth or AGN feedback \citep[see][for details]{Weinberger+18, Pillepich+18a}.
In this work we use the publicly available results from the TNG100-1 run at z=0 (snapshot 99), that consists of a cubic volume with box length of 110.7 Mpc, with dark mater and baryonic mass resolutions being $7.5\times10^6$\Msun and $1.4\times10^6$\Msun, respectively. 
We select all subhalos, identified by means of the \texttt{SUBFIND} algorithm \citep{Springel+05b}, with total stellar mass within two effective radii in the range $10^9<M_*/$\Msun$<10^{12}$.

As discussed in \citet{Corcho-Caballero+21a}, this suite of simulations presents a significantly large number of galaxies that have not formed any stellar particle during the last 2-4 Gyrs.
Most of these systems have experienced strong quenching followed by a sudden gas suppression, remaining red and dead ever since.
Thus, the IllustrisTNG sample comprises an ideal set of SFHs for exploring the properties of both recently-quenched and long-dead galaxies along the AD.

Following a similar approach as with the observational data, we use all the stellar particles confined within \reff{1.5} in terms of total stellar mass.
The integrated star formation histories are thus reconstructed using all the ages and initial masses of each individual particle and then resampled to the same bins as the \texttt{GSD156} library.
We then compute the synthetic stellar continuum spectrum of each galaxy using the \texttt{GSD156} SSP library.

\subsection{\ha nebular emission}
\label{sec:ha_models}
In this work, we assume that the only source of ionizing photons in a galaxy corresponds to the underlying stellar population.
\cchange{In particular, we will assume that the contribution to \ha emission comes entirely from young massive O and B stars \citep[e.g.][]{Kennicutt98, KennicuttEvans12}.}
\cchange{
According to \citet{Kennicutt98}, for solar abundances and a Salpeter IMF (0.1-100 $\Msun$), the connection between the SFR and the \ha luminosity can be well described by
\begin{equation}
SFR ~[\Msun / {\rm yr}] = 7.9 \times 10^{-42} L^{\rm neb}_{\ha} ~[\rm erg\,s^{-1}],
\end{equation}
where we set the characteristic timescale traced by \ha to $20$ Myr (i.e. only stars younger than this age limit will contribute to the ionising photon budget; $SFR\equiv\frac{M_{20}}{20~\rm Myr}$, where $M_{20}$ denotes the stellar mass formed in the last 20 Myr).
}
There might be some particular cases, e.g. \cchange{post-AGB stars}, strong AGN activity or shock/stellar winds, for which \ha emission can be produced by other processes, but on statistical terms, and specially on integrated scales, they should not play such an important role as the radiation emitted by stars \citep{Sanchez+20, Sanchez+21a}.

\section{Results}
\label{sec:results}
Building upon the results obtained in \citetalias{Casado+15} and \citetalias{Corcho-Caballero+21b}, in the present work we aim to provide further insights on the connection between the distribution of nearby galaxies over the AD and their star formation histories.
\cchange{In Section~\ref{sec:the_ageing_diagram} we address the systematic effect, previously ignored, produced by dust extinction on the stellar continuum, by employing the dust-corrected or dust-insensitive proxies \gr and \balbreak, respectively} and establish a
classification of galaxies in terms of their location.
\cchange{In Section~\ref{sec:AD_classification} we show the distribution of characteristic timescales in terms of the adopted demarcation lines for IllustrisTNG galaxies.}
Finally, in Section~\ref{sec:AD_statistics} we study the fraction of galaxies in the Local Universe that undergo secular (ageing) or fast evolution (quenching) comparing the previous results found for IllustrisTNG with those of CALIFA and MaNGA samples.

\subsection{The Ageing Diagram}
\label{sec:the_ageing_diagram}

\begin{figure*}
	\includegraphics[width=\linewidth]{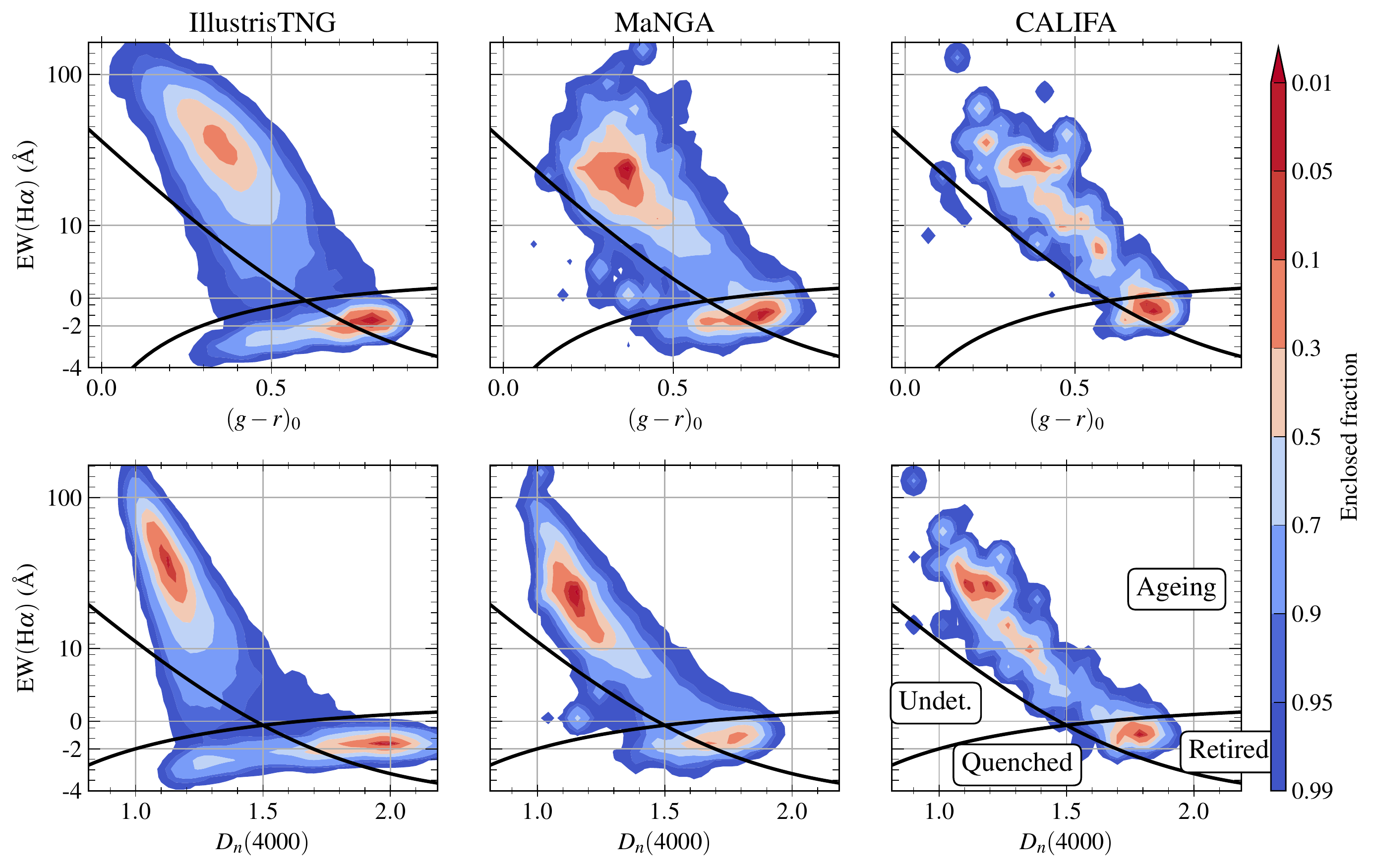}
	\caption{\cchange{Probability distribution of galaxies along the Ageing Diagram for IllustrisTNG (left), MaNGA (middle) and CALIFA (right) samples.
			Contours denote the region encompassing a given fraction of the total sample.
			Solid lines delimit each AD domain: Ageing, Undetermined, Quenched and Retired.}}
	\label{fig:ad_distrib}
\end{figure*}

\cchange{The distribution of galaxies across the ageing diagram for IllustrisTNG, MaNGA and CALIFA samples is shown in Figure \ref{fig:ad_distrib} (left, middle and right panels, respectively).}
The distribution obtained from IllustrisTNG reproduces remarkably well the blue end, corresponding to the so-called star-forming main sequence \citep[e.g.][]{Noeske+07}, as noted previously in \citet{Corcho-Caballero+21a}, albeit we see some slight differences in the red end.
Most notably, the distribution in the \balbreak-\ew diagram seems to reach lower values in the predictions from IllustrisTNG than suggested by the observational samples.
To a lesser extent, we also notice that the colour distribution observed in MaNGA for $\ew<0$ is perhaps slightly broader than predicted from the simulations.
Finally, we find that the passive population of simulated galaxies feature \ew approximately 1 \AA\, below the observed distribution.

\cchange{In \citetalias{Corcho-Caballero+21b}, we provided a division of the AD in terms of four different domains, namely: \textit{blue-emission} -- \ew$>0$, $g-r<0.7$ --, \textit{red-emission} -- \ew$>0$, $g-r\geq0.7$ --, \textit{blue-absorption} -- \ew$\leq0$, $g-r<0.7$ -- and \textit{red-absorption} -- \ew$\leq0$, $g-r\geq0.7$.
In the present work we aim to improve this classification and explore whether the location of galaxies in the ageing diagram is able to effectively discriminate among different evolutionary pathways.}

\cchange{We rely on the results obtained from both observational and numerical samples, to constrain the preferred location of ageing and quenched galaxies in the AD, respectively.}
According to these results, we provide \ccchange{phenomenological} demarcation lines that divide the AD into four different regimes using the following expression:
\begin{equation}
\label{eq:ad_fit}
\rm EW(x) = k \times 10^{-\alpha x} + EW_\infty
\end{equation}
where $x$ denotes either $(g-r)_0$ or $D_n(4000)$, and $\rm EW_\infty$ corresponds to the asymptotic equivalent width of the \ha absorption line reached by the oldest stellar populations.
As shown in Figure \ref{fig:tng_AD}, galaxies in the ageing diagram are now distributed across four domains, delimited by the two solid black lines, $\rm EW(H\alpha)_{ageing}(x)$ \cchange{(roughly delimiting the distribution with $\ew>0$)} and $\rm EW(H\alpha)_{quench}(x)$, \cchange{(encompassing systems with $\ew$ in absorption)} whose adopted parameter values are shown on Table~\ref{tab:best_params}.
\ccchange{The proposed parameters are arbitrarily chosen such that the quenched line comprises the sequence of galaxies with values of \ew in absorption, clearly seen in both observed and simulated data.
The ageing line, on the other hand, is meant to encompass the lower envelope of the population of galaxies with \ew in emission (roughly corresponding to the contour containing 90 per cent of the distribution), and the extrapolation towards the red end is guided by the results reported in Figure~\ref{fig:tng_AD} (see the discussion in Section~\ref{sec:AD_classification} below)}.
\textit{Ageing} galaxies (AGs) will lay above both demarcation lines, while \textit{quenched} galaxies (QGs) will be located below the two of them.
Below the ageing and above the quenching lines, we find an \textit{undetermined} population (UGs) that might be made up of quenching galaxies and/or rejuvenated systems, whose current evolutionary status is very difficult to address from an observational point of view.
Finally, the \textit{retired} domain contains all objects above and below the ageing and quenched sequence, respectively.
Note that galaxies can end up in the retired sequence either through quenching or ageing evolutionary paths, although the retired population in IllustrisTNG is strongly dominated by systems suddenly quenched several Gyr ago.
Once galaxies reach the retired regime, it becomes extremely challenging to discriminate their past SFH by means of the AD.

\begin{table}
	\centering
	\begin{tabular}{c|c|c|c}
		\hline
		Line & k (\AA) & $\alpha$ & $\rm EW_\infty$ (\AA)\\
		\hline
		${\rm EW(H\alpha)_{ageing}}((g-r)_0)$ & 45.0 & 1.7 & -4.50 		\\
		${\rm EW(H\alpha)_{quench}}((g-r)_0)$ & -7.0 & 0.9 & 1.80 		\\
		${\rm EW(H\alpha)_{ageing}}(D_n(4000))$ & 3915.0 & 2.3 & -2.00	\\
		${\rm EW(H\alpha)_{quench}}(D_n(4000))$ & -5.9 & 0.5 & 0.05	\\
	\end{tabular}
	\caption{AD diagram demarcation lines adopted values of eq.~\eqref{eq:ad_fit} for the \gr-\ew and \balbreak-\ew versions.}
	\label{tab:best_params}
\end{table}

\subsection{Ageing, Quenched and Retired Galaxies in IllustrisTNG}
\label{sec:AD_classification}
\begin{figure}
	\includegraphics[width=\linewidth]{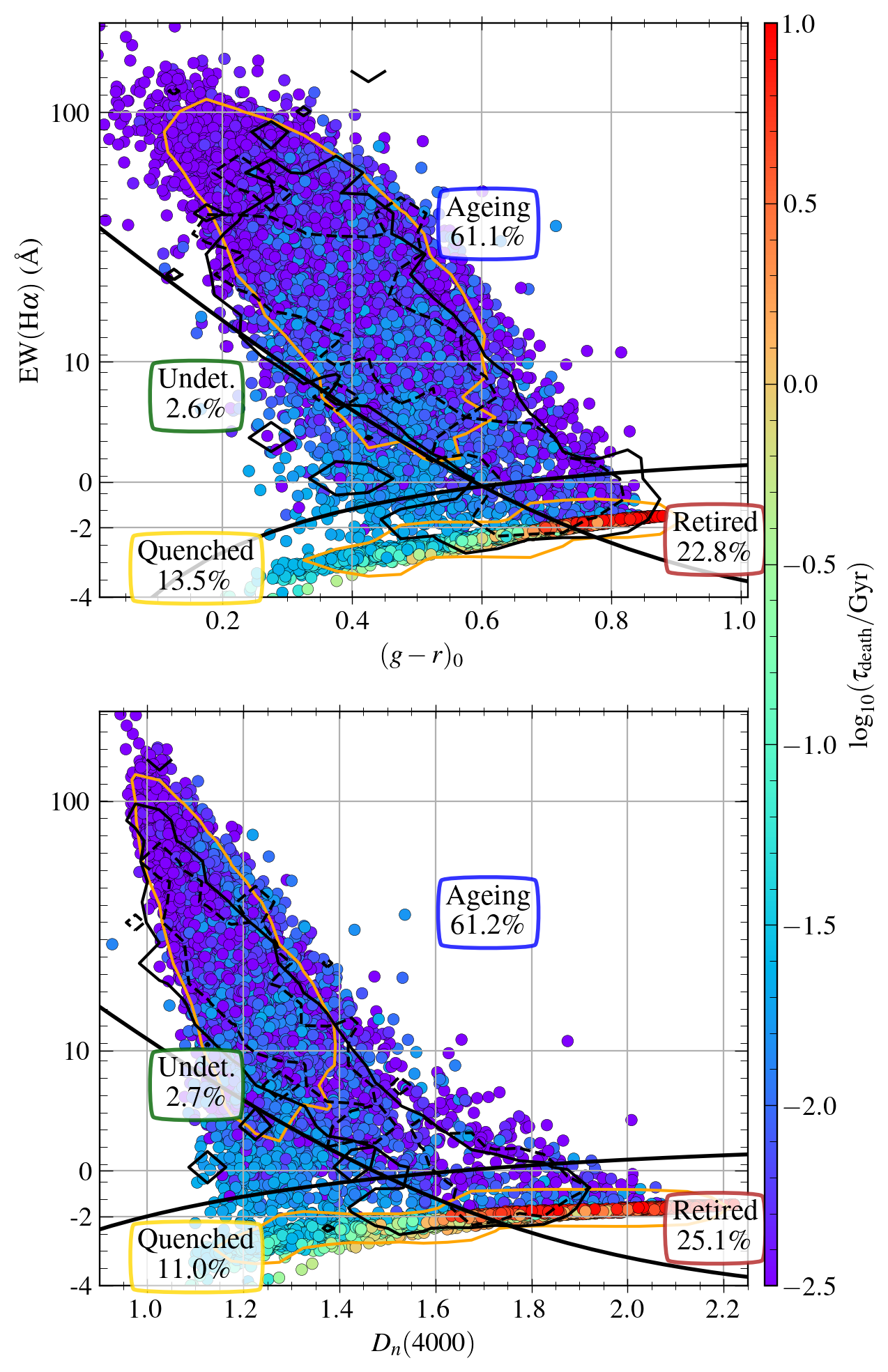}
	\caption{Ageing diagram for IllustrisTNG galaxies, coloured by \taudeath.
	Top and bottom panels show the $(g-r)_0$ and $D_n(4000)$ versions of the AD, respectively.
	Left column uses the SF model for \ha emission while the SF-PL is used for the right column.
	Solid lines denote the limits for each AD domain: Ageing, Undetermined, Quenched and Retired.
	The fraction of galaxies within each domain is shown below the label.
	\cchange{Orange and black contours enclose 90 per cent of the IllustrisTNG and MaNGA (solid) and CALIFA (dashed) samples, respectively.}}
	\label{fig:tng_AD}
\end{figure}

We rely on the results obtained from IllustrisTNG to \cchange{interpret} the preferred location of ageing and quenched galaxies in the AD, respectively.
In particular, we use of the amount of time since star formation completely ceased, referred to as death time, \taudeath, to discriminate between ageing galaxies ($\taudeath \lesssim \cchange{20}$~Myr) and systems that have been quenched between $\taudeath \sim \cchange{20} $~Myr and a few Gyr ago.
Numerically, \taudeath corresponds to the age of the youngest stellar particle linked to each galaxy, under the constrains outlined in~\ref{sec:IllustrisTNG}.
A discussion of resolution effects can be found in Section~\ref{sec:discussion} and appendix~\ref{appendix:timescales}.

Figure \ref{fig:tng_AD} shows the distribution of IllustrisTNG galaxies across the AD, coloured by their corresponding \taudeath.
We also include the contour that encompasses the 90 per cent of the total sample \cchange{for IllustrisTNG, MaNGA and CALIFA,} to illustrate the location of the bulk of the distribution.
Purple points ($\taudeath \lesssim \cchange{20}$~Myr) will be considered systems whose star formation histories show recent changes merely due to its intrinsic stochastic nature, arranging thus along the \emph{ageing sequence}.
This population accounts for the majority of galaxies in the sample and contains a wide range of different evolutionary stages: from blue stellar continuum and high \ew (galaxies in the MS) to red colours and practically null \ha emission (quiescent population).
In addition, we find \cchange{that the} detached population of galaxies arranging over a clearly separated location of the AD, referred to as the \emph{quenched sequence}, \cchange{is} composed of systems with recently truncated SFHs.
The death times range from \cchange{tens} of Myr, for galaxies still displaying blue colours and strong stellar continuum features, with prominent \ha absorption and low \balbreak (akin to post-starburst systems), up to red-and-dead galaxies that have not formed a single stellar particle during the last $\gtrsim 3$ Gyr (\cchange{some of which} can be interpreted as the descendants of PSBGs).
\change{The minimum \taudeath in this domain} will be close to the age limit adopted for young stellar populations in our \ha emission model, $\tau_{\rm young}$, \change{that sets the minimum transition timescale from the ageing to the quenching sequence in our models}.
The typical values of \taudeath for quenched systems suggest that the quenching timescales, i.e. the duration of the transition \change{through} the quenched sequence probed by this diagram, corresponds to \change{$\sim 1$~Gyr}.
Galaxies whose star formation activity is interrupted on shorter timescales (fast quenching) will be found in the quenched sequence.
On the other hand, systems experiencing slow quenching processes will appear as ageing galaxies.

\begin{figure}
	\includegraphics[width=\linewidth]{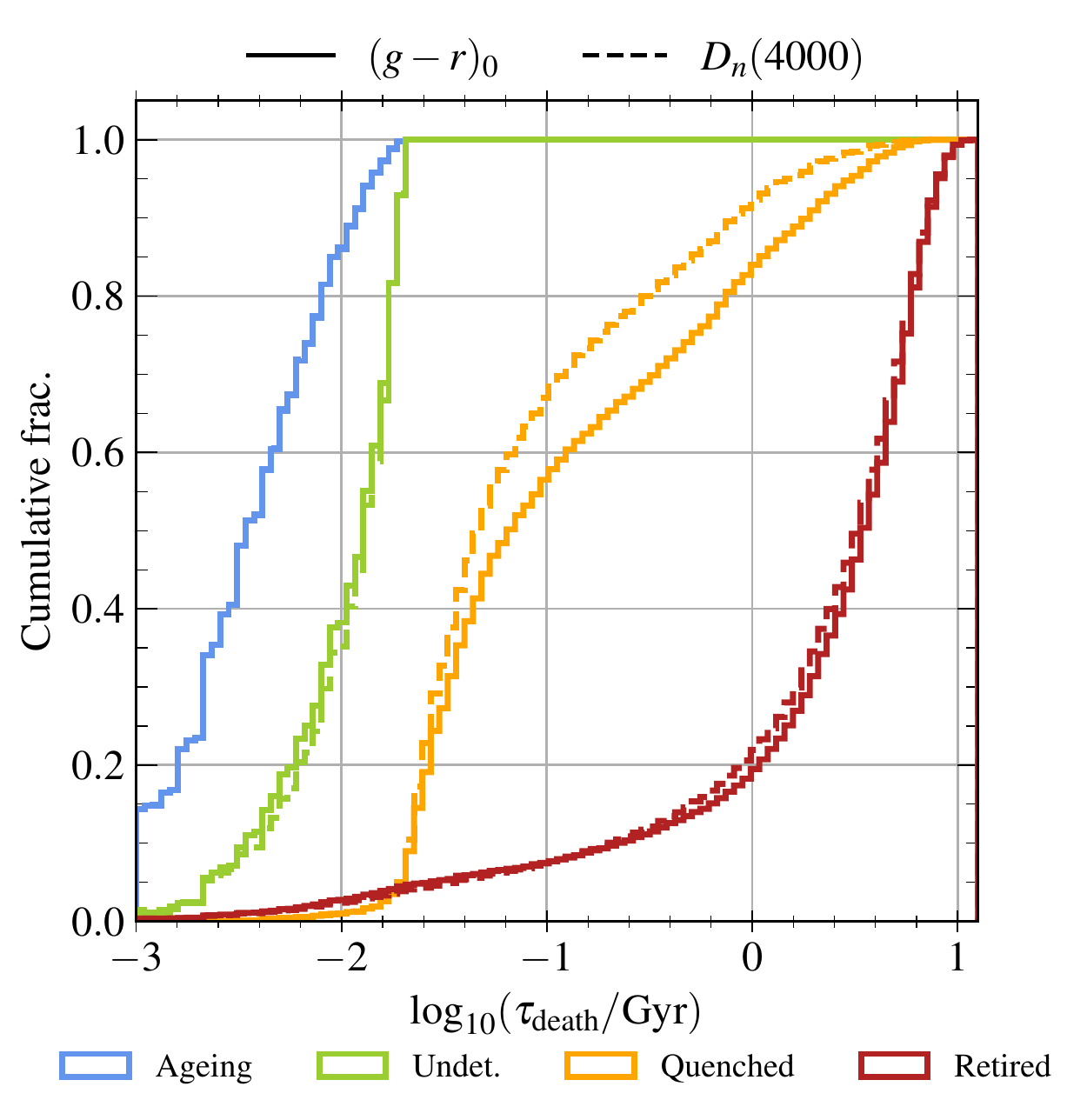}
	\caption{Cumulative distribution of death times \taudeath in terms of the AD domains -- Ageing, Undetermined, Quenched and Retired - \cchange{using both \gr (solid) and \balbreak (dashed) for IllustrisTNG galaxies}.}
	\label{fig:death_time_fractions}
\end{figure}

\begin{table}
	\centering
	\begin{tabular}{c|c|c|c|c}
		\hline

		$\log_{10}(\frac{\taudeath}{\rm Gyr})$ & Ageing & Undet. & Quenched & Retired \\
		\hline
		\gr & $-2.46^{+0.41}_{-0.41}$ & $-1.90^{+0.16}_{-0.43}$ &$-1.18^{+1.19}_{-0.43}$ & $0.54^{+0.27}_{-0.68}$ \\
		\hline
		\balbreak & $-2.46^{+0.41}_{-0.40}$ & $-1.89^{+0.15}_{-0.39}$&  $-1.33^{+0.99}_{-0.30}$ & $0.51^{+0.29}_{-0.74}$ \\

	\end{tabular}
	\caption{Median \taudeath and dispersion (computed using the 16 and 84 percentiles) of galaxies within each domain of the ageing diagram for IllustrisTNG galaxies.}
	\label{tab:tau_death}
\end{table}

Following the proposed classification, we show the cumulative distribution of \taudeath within each regime in Figure~\ref{fig:death_time_fractions}.
\cchange{The median and 16, 84 percentiles} of \taudeath for each galaxy domain \cchange{are quoted on} Table~\ref{tab:tau_death}.
The vast majority (\cchange{$61\%$}) of IllustrisTNG galaxies are classified as ageing, and they feature SFHs compatible with smooth secular evolution up to the present epoch, whereas the retired population adds up to \cchange{$23-25$} per cent of the total fraction of galaxies.
On the other hand, less than \cchange{$\sim3\%$} of galaxies qualify as undetermined (probably in the process of quenching), while quenched systems account for \cchange{11 to 14} per cent of the total sample.

\subsection{Ageing and Quenching in the Local Universe}
\label{sec:AD_statistics}
\begin{figure*}
	\includegraphics[width=\linewidth]{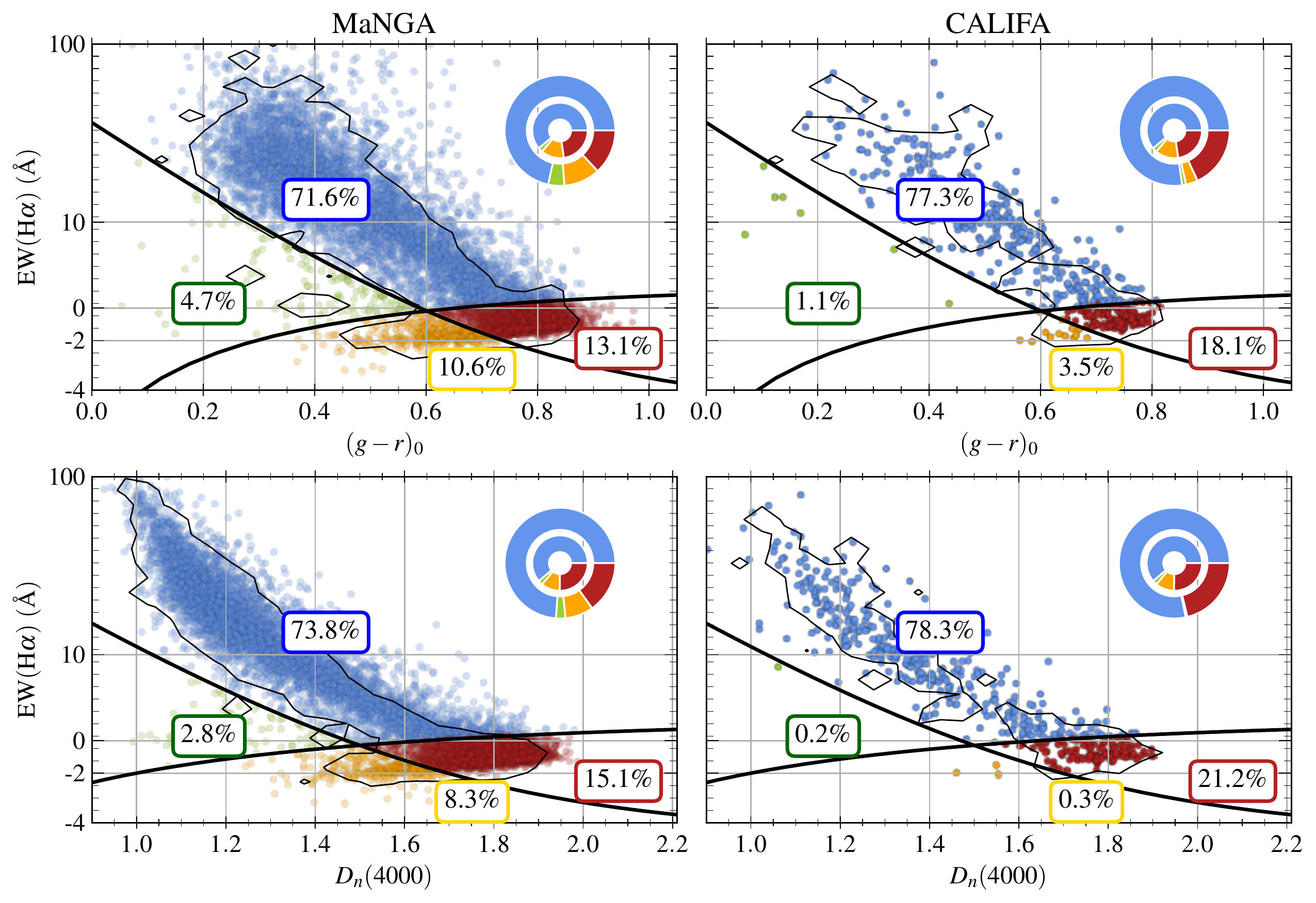}
	\caption{Ageing diagram for MaNGA and CALIFA samples computed using the empirical measurements.
	Galaxies are coloured according to the classification provided in Section~\ref{sec:AD_classification}. On each panel the fraction of galaxies within each class is shown as a pie chart (outer ring), with the results obtained from IllustrisTNG (inner ring) for comparison. Black contours denote the 90 per cent of the total density distribution.}
	\label{fig:AD_observational}
\end{figure*}

Based on the characterization of the AD in terms of four domains explored the previous section, we can now investigate the fraction of galaxies that populate each AD domain using CALIFA and MaNGA samples, in order to address the importance of quenching in the Local Universe.
To perform this analysis we use the empirical measurements of \ew, rather than the predicted values derived from the model.

On Figure~\ref{fig:AD_observational} we show the distribution of galaxies across the two versions of the AD for the MaNGA (left) and CALIFA (right) samples.
Each galaxy has been coloured according to the domain they belong to: ageing (blue), undetermined (green), quenched (yellow) and retired (red).
In MaNGA, we provide some transparency to the data points in order to emphasize where the bulk of the distribution is located, and a black contour is included at every panel containing the 90 per cent of the total density distribution.
Additionally, we include on the top-right corner of each panel a pie chart, using the same colour coding as the data points, that represents the volume-corrected fraction of galaxies found on each domain (outer ring), compared with the results found for IllustrisTNG (inner ring).

We find that there is overall a very good agreement between IllustrisTNG domain fractions and the results obtained from the observational data.
Yet, the fraction of galaxies in the ageing domain seems to be slightly higher in the latter, while the fraction of retired systems is systematically larger for the cosmological simulation.
At variance with the results found for CALIFA, the distribution of MaNGA galaxies shows fairly high fractions of undetermined ($3-5\%$) and quenched ($8-10\%$) systems.
In the CALIFA sample we find a clear dominance of ageing galaxies ($\sim77-78\%$), and a negligible number of quenched systems ($\lesssim 3\%$).
However, the number of galaxies in the CALIFA sample is significantly lower, precluding from any robust conclusion regarding the quenched population.
\cchange{This is} in agreement with previous results found in \citetalias{Corcho-Caballero+21b}\cchange{, where it was shown that recently quenched galaxies are usually dwarf systems with early-type morphologies; highly under-represented in the CALIFA Mother Sample, where completeness is only achieved above $\log_{10}(M_*/\rm M_\odot) \gtrsim 9.65$ \citep{CALIFAselection}.
A more thorough analysis of the physical properties of Ageing, Undetermined, Quenched and Retired populations will be presented in a forthcoming work (Corcho-Caballero et al. in prep.)}.

\section{Discussion}
\label{sec:discussion}

One of the most interesting aspects of the ageing diagram
lies on its ability to trace recent changes in the SFH of galaxies, beyond their present evolutionary status (e.g. star-forming vs quiescent/passive).

In this section, we will discuss the interpretation of the AD classification in terms of the physical properties of the galaxies.
In Section~\ref{sec:star_formation_histories} we explore the typical star formation history of galaxies within each domain.
Then, we investigate the time spent at different star formation activity levels in Section~\ref{sec:timescales}, in connection with the timescale of the transition from the blue cloud to the red sequence.
In Section~\ref{sec:mass_formed}, we propose an almost one-to-one correspondence between the AD classification and the stellar mass formed in the last $\cchange{\tshort}$~Myr and $3$~Gyr.
Finally, we discuss in Section~\ref{sec:sequences} the implications of the location along the ageing and quenched sequences in the context of galactic evolution.

\subsection{Star formation histories}
\label{sec:star_formation_histories}
\begin{figure*}
	\includegraphics[width=\linewidth]{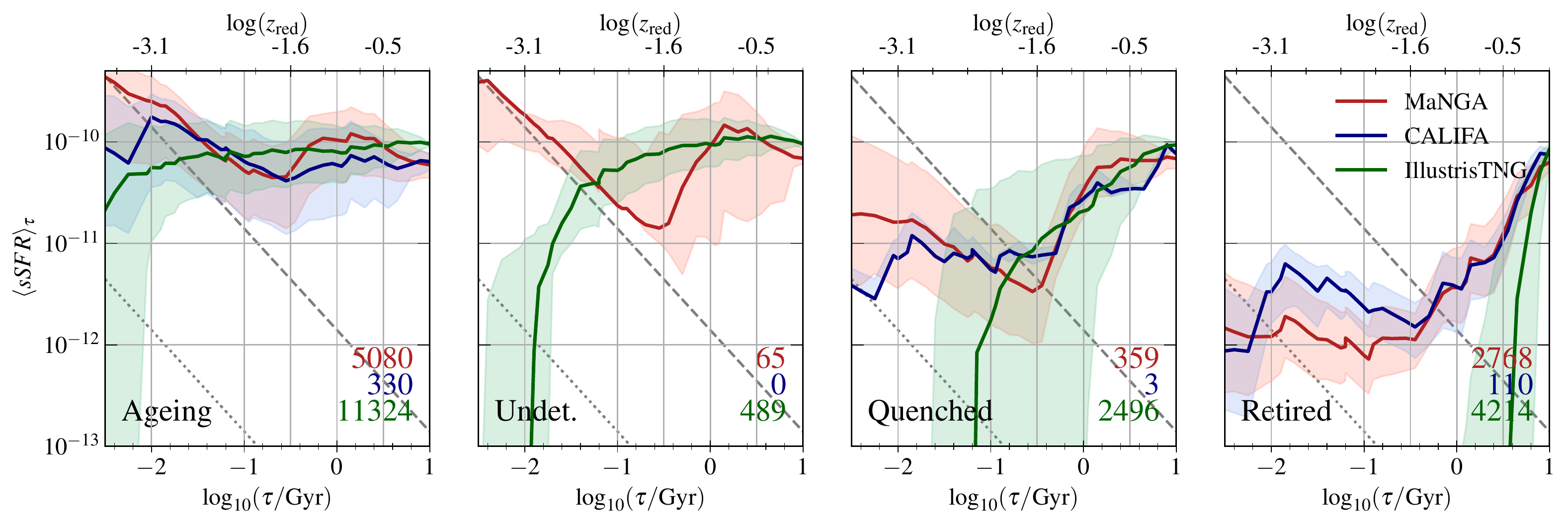}
	\caption{50 (solid line), 84 and 16 percentiles (shaded coloured region) of MaNGA/CALIFA \pipe (red/blue) and IllustrisTNG (green) average star formation histories as function of lookback time and redshift. Panels correspond to each AD domain, with the number of galaxies used to computed the percentiles shown on the lower-right corner.
	Grey \ccchange{dashed and dotted lines} denote the estimates of IllustrisTNG100 $sSFR$ resolution limits for a galaxy with total stellar mass of $10^9\,\rm M_\odot$ and $10^{11}\,\rm M_\odot$, respectively.}
	\label{fig:ssfr_histories}
\end{figure*}

The ageing diagram shows that there are at least two evolutionary pathways, with most of the galaxies evolving on time scales comparable to the age of the Universe, while a reduced number of systems ($1-10\%$) undergo some quenching process that truncates their star formation in less than a few hundred Myr.
Here we explore the main trends of the SFHs within each of the four AD domains by computing the average specific star formation rate during the last $\tau$ Gyr:
\begin{equation}
\langle sSFR \rangle(\tau) \equiv \frac{1}{M^{\rm tot}_*}\frac{\sum_i M_i(\tau_i < \tau)}{\tau}
\end{equation}
over the SSP ages and stellar particles in \pipe and IllustrisTNG, respectively.

In Figure~\ref{fig:ssfr_histories} we show the volume-weighted median values and dispersion (16 and 84 percentiles) of $\langle sSFR\rangle (\tau)$ for MaNGA (red), CALIFA (blue) and IllustrisTNG (green) galaxies within each AD domain.
We have required that both diagrams -- \ew-\balbreak and \ew-\gr -- provide the same class; otherwise the galaxy will be labelled as \emph{unclassified} and will not be considered in the estimation.

There is a clear trend of decreasing $sSFR$ activity from ageing systems to retired galaxies.
Galaxies within the Ageing domain present roughly constant star formation histories, with mean $sSFR$ ranging from $\sim10^{-9.5}~\rm yr^{-1}$ to $\sim10^{-11}~\rm yr^{-1}$, suggesting that this has been their evolutionary status throughout cosmic history.
On the other hand, galaxies classified as undetermined or quenched are found to display steep drops in their SFH during the last $\sim$~Gyr.
In particular, quenched galaxies appear to have reduced their SF activity by at least one order of magnitude during the last Gyr.
In CALIFA and MaNGA, they still present mild levels of star formation (around $\sim 10^{-11}$~yr$^{-1}$), whereas in IllustrisTNG the star formation activity ceases completely.
Finally, the vast majority of Retired systems in the simulation show an abrupt interruption of the SF activity about $\sim 3$~Gyr ago ($z\sim 0.3$), with no subsequent star formation, while the decline seems to be somewhat more gradual for both observational samples, who also feature residual star formation with $sSFR\sim 10^{-12}~\rm yr^{-1}$.

As recently discussed in the literature \citep[e.g.][]{Katsianis+21, Corcho-Caballero+21a}, galaxies in cosmological simulations such as IllustrisTNG-100 are not able to retain low levels of SF (possibly due to a strong feedback implementation), while observational estimates provide a different picture where red-and-dead galaxies still form stars, albeit at a much lower rate than during the early phases of their evolution.
To assess whether this could be an effect of the finite resolution (particle mass) in the simulations, we plot \ccchange{in Figure~\ref{fig:ssfr_histories}} the maximum $sSFR$ achievable, corresponding to one single baryonic particle during the last $\tau$ Gyr, for a galaxy with a total stellar mass of $10^9~\rm M_\odot$ (dashed) and $10^{11}~\rm M_\odot$ (dotted) \citep[see also appendix A in][]{Donnari+19}.
Although resolution effects are important below a few hundred Myr (i.e. Quenched galaxies), and we find a good qualitative agreement with the empirical estimates at almost every panel above the dashed line, the truncation of the SFH in Retired systems is well resolved in the simulation, and we argue that the discrepancy found above the dashed line cannot be attributed to resolution effects.
On the other hand, the reconstruction of low levels of star formation is far from straightforward \citep[e.g][]{Salvador-Rusinol+20}, and it would also be possible that \pipe SFHs below $\sim 10^{-11}$~yr$^{-1}$ ought to be considered as upper limits.

Regardless of this issue, which is of the utmost importance both from an observational and a theoretical perspective, we conclude that the AD classification corresponds to physically meaningful classes in terms of the recent SFH.
Ageing systems are representative of the so-called main sequence of star formation, while Quenched galaxies have undergone a sudden drop on short time scales.
For Retired galaxies, the AD is unable to tell whether star formation has steadily dropped over cosmic time, or galaxies died (completely shut off star formation) a long time ago.

\subsection{Ageing and quenching timescales}
\label{sec:timescales}
\begin{figure*}
	\includegraphics[width=\linewidth]{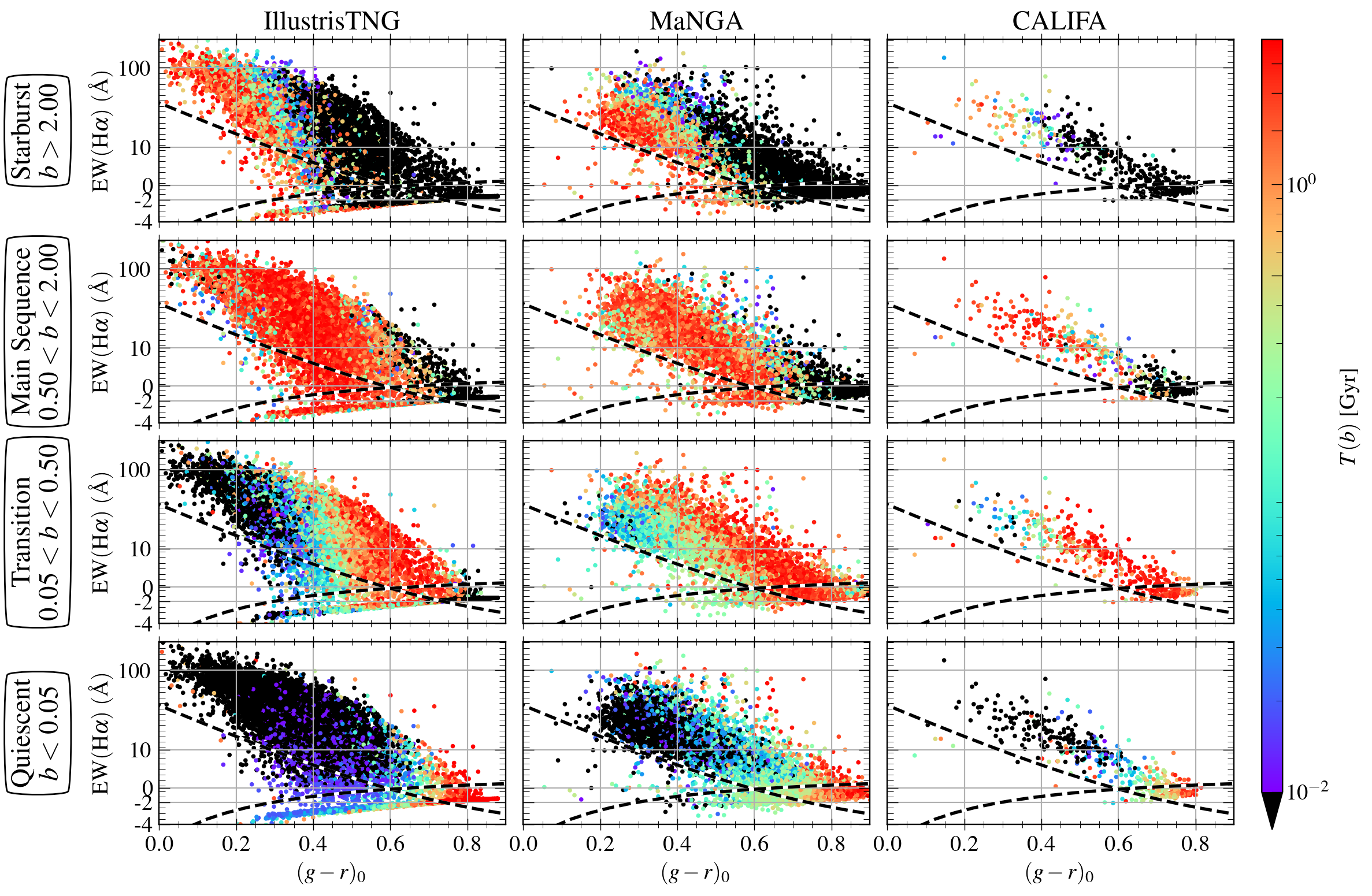}
	\caption{Distribution of time spent ($T$) at each SF activity regime (SB, MS, T, Q) across the \gr-\ew ageing diagram, during the last 3 Gyr.
    Each column corresponds to IllustrisTNG (SF and SF+PL models), MaNGA and CALIFA samples, respectively.
	Dashed black lines delimit the four AD domains described in Section~\ref{sec:AD_classification}.}
	\label{fig:ad_timescales}
\end{figure*}

\begin{table}
	\centering
	\begin{tabular}{c|c|c|c}
		\hline Ageing & \cchange{IllustrisTNG} & MaNGA & CALIFA\\ \hline 
		\textit{Starburst} & $0.05^{+1.08}_{-0.05}$ & $0.53^{+1.49}_{-0.50}$ & $0.06^{+0.79}_{-0.06}$\\[0.2cm] 
		\textit{Main-Sequence} & $2.29^{+0.62}_{-1.22}$ &  $1.43^{+0.84}_{-0.92}$ & $1.77^{+0.55}_{-1.44}$\\[0.2cm] 
		\textit{Green-Valley} & $0.08^{+1.03}_{-0.06}$ &  $0.29^{+1.23}_{-0.22}$ & $0.69^{+1.74}_{-0.63}$\\[0.2cm] 
		\textit{Quiescent} & $0.00^{+0.04}_{-0.00}$ &  $0.02^{+0.07}_{-0.02}$ & $0.00^{+0.06}_{-0.00}$\\[0.2cm] 
		\hline Undet. & IllustrisTNG & MaNGA & CALIFA\\ \hline 
		\textit{Starburst} & $0.17^{+1.31}_{-0.17}$ &  $0.85^{+0.67}_{-0.83}$ & NaN\\[0.2cm] 
		\textit{Main-Sequence} & $2.38^{+0.49}_{-1.16}$ &  $0.39^{+1.16}_{-0.23}$ & NaN\\[0.2cm] 
		\textit{Green-Valley} & $0.08^{+0.40}_{-0.06}$ &  $0.29^{+2.13}_{-0.24}$ & NaN\\[0.2cm] 
		\textit{Quiescent} & $0.03^{+0.02}_{-0.02}$ &  $0.15^{+0.21}_{-0.15}$ & NaN\\[0.2cm] 
		\hline Quenched & IllustrisTNG & MaNGA & CALIFA\\ \hline 
		\textit{Starburst} & $0.00^{+0.65}_{-0.00}$ &  $0.06^{+0.83}_{-0.06}$ & $0.00^{+0.00}_{-0.00}$\\[0.2cm] 
		\textit{Main-Sequence} & $1.49^{+1.06}_{-1.29}$ &  $1.69^{+0.54}_{-1.11}$ & $0.93^{+0.41}_{-0.11}$\\[0.2cm] 
		\textit{Green-Valley} & $0.64^{+1.53}_{-0.55}$ &  $0.42^{+1.40}_{-0.18}$ & $1.30^{+0.56}_{-0.65}$\\[0.2cm] 
		\textit{Quiescent} & $0.08^{+0.58}_{-0.05}$ &  $0.27^{+0.09}_{-0.13}$ & $0.04^{+0.05}_{-0.03}$\\[0.2cm] 
		\hline Retired & IllustrisTNG & MaNGA & CALIFA\\ \hline 
		\textit{Starburst} & $0.00^{+0.00}_{-0.00}$ &  $0.00^{+0.00}_{-0.00}$ & $0.00^{+0.00}_{-0.00}$\\[0.2cm] 
		\textit{Main-Sequence} & $0.00^{+0.00}_{-0.00}$ & $0.00^{+0.39}_{-0.00}$ & $0.00^{+0.05}_{-0.00}$\\[0.2cm] 
		\textit{Green-Valley} & $0.00^{+0.86}_{-0.00}$ & $1.93^{+0.53}_{-0.88}$ & $2.12^{+0.38}_{-0.90}$\\[0.2cm] 
		\textit{Quiescent} & $2.96^{+0.03}_{-1.04}$ & $0.75^{+1.14}_{-0.38}$ & $0.70^{+1.02}_{-0.25}$\\[0.2cm] 
	\end{tabular}
	\caption {Median, 16th and 84th percentiles of time spent (Gyr) at each star-forming activity regime for galaxies classified in terms of the AD domains. }
	\label{tab:timescales}
\end{table}

Understanding the characteristic timescales that drive galaxy evolution is extremely challenging.
Observationally, the choice of proxies used for tracing changes on the SFH of a galaxy -- morphological features, chemical composition or different SFR indicators -- will have a net impact on the resulting timescales inferred \citep[e.g.][]{Wetzel+13, Peng+15}.
Also, the different thresholds adopted for establishing discrete and distinct evolutionary stages (e.g. `main sequence', `green valley', `quenched') may lead to a wide range of results \citep{Wright+19, Tacchella+22, Walters+22}.
In particular, galaxy quenching has been typically studied by means of the time spent by galaxies in crossing some arbitrary boundaries, such as the Green Valley in the colour-magnitude diagram, or the transition from the main sequence of star-forming galaxies and the passive population in terms of $SFR$ \citep[e.g.][]{Schawinski+14, Wright+19, Walters+22}.

In this work we will follow the latter approach and explore the typical timescales that drive changes on the star formation history of galaxies by means of the birth-rate parameter\footnote{
\ccchange{
Note that, neglecting stellar mass loss, $M_*(t) \approx \int_0^t SFR(t) \dd t$ would be the total stellar mass formed, and $b(t) \approx t\ sSFR(t)$.
}} \citep{Kennicutt83}
\begin{equation}
\ccchange{
b(t) = \frac{SFR(t)}{\langle SFR \rangle} = \frac{t\ SFR(t)}{\int_0^t SFR(t) \ {\rm d}t}
},
\end{equation}
where $t$ denotes cosmic time,
and establish four regimes of star formation that represent different stages of galaxy evolution:

On the one hand, we refer to star-forming galaxies with $0.5<b<2$ as `\textit{Main Sequence}' (MS).
This choice is motivated by the fact that $b=0.5$ corresponds to the regime at which the galaxy is 0.3 dex below the mean $sSFR$, similar to the observed scatter \citep[e.g.][]{Noeske+07}.
Analogously, we use the term `\textit{Starburst}' (SB) to denote systems above $b > 2$.

On the other hand, we define a galaxy as `\textit{Quiescent}' (Q) if $b < 0.05$, i.e. an order of magnitude below the lower limit of the main sequence at time $t$.
Therefore, the range $0.05 < b < 0.5$ denotes the `\textit{Transition}' (T) regime where galaxies form stars at an intermediate rate.
In this paper we interpret the time a galaxy spends in this regime as the timescale to move from the Main Sequence to the Quiescent evolutionary state.

Let us note that this definition bears some relevant nuances with respect to other alternatives used in the literature.
For the present time $t \simeq 13.7$~Gyr, $b = 0.5$ and $0.05$ correspond to $sSFR \simeq 3 \times 10^{-11}$ and $3 \times 10^{-12}$~yr$^{-1}$, respectively.
Overall, these boundaries are roughly correlated with the observed colour distribution: Ageing galaxies currently in the Main Sequence, Transition, and Quiescent state will tend to display blue, green, and red colours, respectively.
However, a galaxy undergoing fast quenching will quickly move from MS to Q (i.e. spend a short time in the Transition stage), retaining almost its original colour throughout the process.
Moreover, a constant star formation rate, roughly representative of a Main Sequence galaxy, yields $b = 1$ at all times and a steadily decreasing $sSFR \propto 1/t$.
At a future time $t \sim 100$~Gyr, this system will feature green colours, that will eventually become red when its $sSFR$ drops below $10^{-12}$~yr$^{-1}$ at $t \sim 1000$~Gyr.
However, according to our definition, this galaxy will always be classified as MS, since $b=1$.

For each galaxy we compute the total amount of time $T(b)$ spent within each SF activity regime, based on IllustrisTNG and \pipe SFHs.
If the system goes under several rejuvenation or quenching episodes, they will be combined, providing an `effective' timescale larger than the time required to truncate (or reignite) star formation in each individual event.
In order to mitigate this issue, we have restricted our estimates of the four times $T(b)$ to the last 3 Gyr (i.e. $T_{\rm SB} + T_{\rm MS} + T_{\rm T} + T_{\rm Q} = 3$~Gyr).
The distribution of individual timescales across the extinction-corrected colour-equivalent width ageing diagram is shown in Figure \ref{fig:ad_timescales} for IllustrisTNG and \pipe (MaNGA and CALIFA) star formation histories (see Figure~\ref{fig:ad_timescales_d4000} for the \balbreak version).

We find consistent results between theoretical and empirical estimates of $T$ over each activity regime.
Galaxies on the ageing sequence display a smooth gradual evolution, strongly correlated with their current location within the Ageing domain of the AD.
While systems on the blue-most edge show signatures of starburst activity (in some cases during the whole period of time, $T_{\rm SB} \sim 3$~Gyr), the vast majority of currently blue galaxies have spent most of the time on the MS.
Redder Ageing systems have been mostly forming stars in the Transition regime during the last 3 Gyr, with a negligible fraction spending most of the time in the Quiescent phase (usually, $T_{\rm Q} < 500$~Myr).
On the other hand, those galaxies classified as Retired in terms of the AD have barely spent any time on the SB and MS regimes during the last 3~Gyr, in favour of the T and Q phases.
This suggest that few (if any) galaxies reaching the retired regime in the AD are able to recover star formation rates in the MS (i.e. bursting/gasping star formation).
Finally, Quenched (and, to some extent, Undetermined) galaxies appear as a detached population that has spent a significant amount of time in the MS ($T_{\rm MS} \sim 1-1.5$~Gyr) and crossed the Transition stage on short timescales ($T_{\rm T} \lesssim 1$~Gyr).
The blue end of the quenched sequence shows signatures of recent Starburst activity; most likely, they correspond to the so-called post-starburst galaxies.

Our results are in broad agreement with previous values of quenching timescales reported in the literature \citep{Wetzel+13, Hahn+17, Tacchella+22}.
Moreover, we also find that the distributions of the times $T(b)$ are fairly consistent between IllustrisTNG and the observational samples.
The most significant difference is, once again, that \pipe SFHs display an asymptotic $sSFR$ that places them in the Transition regime for a large fraction of time, whereas simulated galaxies are completely Quiescent ($b < 0.05$).
We provide on Table~\ref{tab:timescales} the median and dispersion values of $T$ for each star formtion phase (SB, MS, T, and Q) in terms of the AD domains (Ageing, Undetermined, Quenched, and Retired) for IllustrisTNG (SF, SF+PL), MaNGA and CALIFA samples.
See also Appendix~\ref{appendix:timescales} (Fig.~\ref{fig:T_distrib}) for a more detailed discussion on the statistical distribution of timescales along the AD and the systematic effects of numerical resolution.

\subsection{Physical interpretation of the AD}
\label{sec:mass_formed}

\begin{figure*}
	\includegraphics[width=\linewidth]{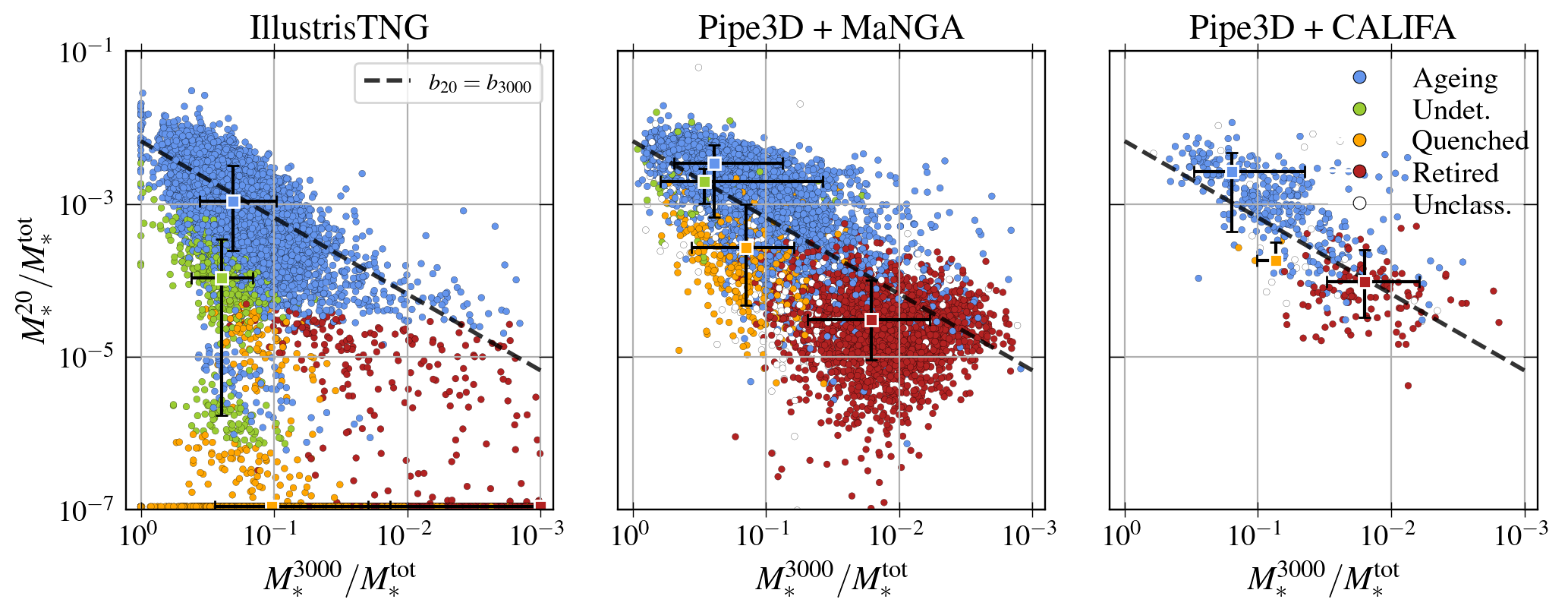}
	\caption{Stellar mass fraction formed during the \cchange{last \tlong~Myr (\masslong) versus the fraction formed in the last \tshort~Myr (\massshort)}.
	Left, central and right panels show the distribution of IllustrisTNG, \pipe+MaNGA and \pipe+CALIFA samples, respectively.
	Each galaxy has been coloured according to its AD class (Ageing, Undetermined, Quenched, Retired) using both versions of the AD. Objects with more than one class are denoted as unclassified and appear as white points.
	Coloured square points represent the median value of each AD class, with error bars denoting the dispersion, computed using the 10 and 90 percentiles, respectively.
	The black dashed line illustrates the constant birthrate scenario.
}
	\label{fig:mass_frac_scatter}
\end{figure*}

\begin{figure*}
	\includegraphics[width=\linewidth]{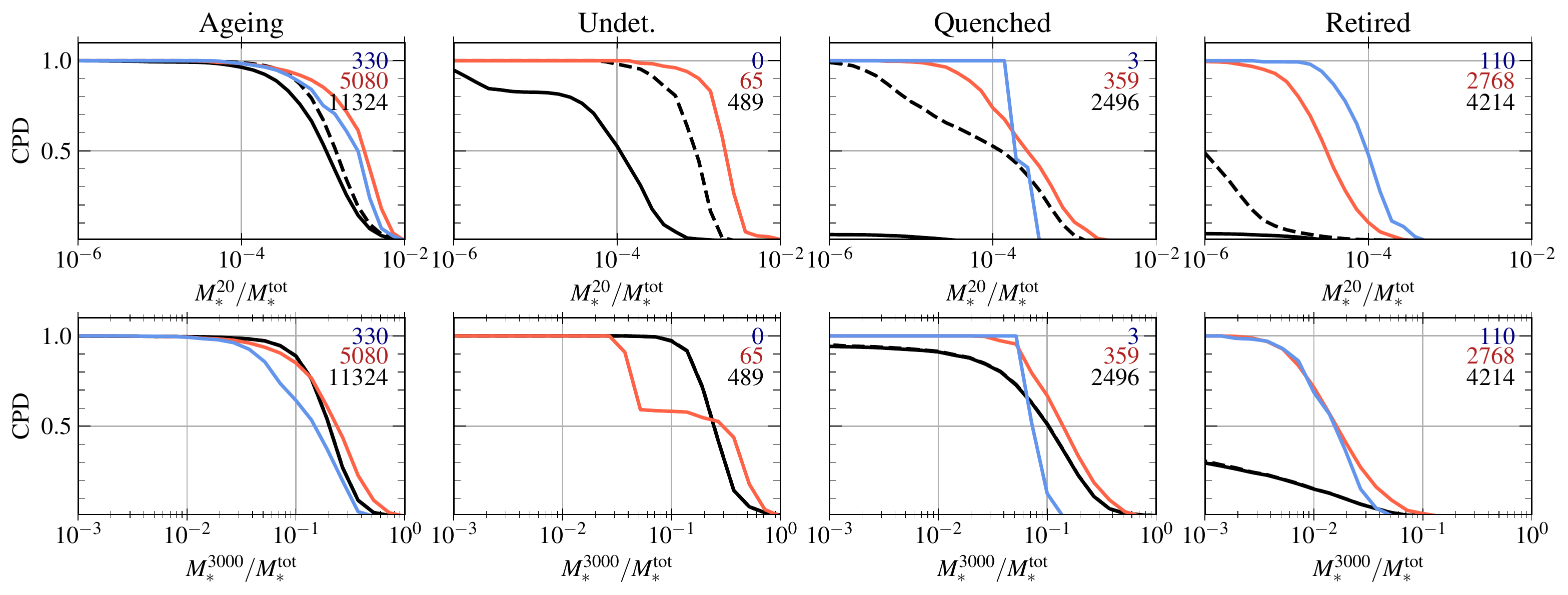}
	\caption{Cumulative probability distribution of the mass fraction formed during the last \cchange{\tshort (top) and \tlong (bottom) Myr}.
	IllustrisTNG distributions are denoted as black solid lines, with dashed lines representing the distribution computed including the formation of an additional stellar particle distributed during the period since star formation stopped.
	MaNGA and CALIFA \pipe distributions are illustrated as red and blue lines.
	Each column denotes an AD domain (Ageing, Undetermined, Quenched and Retired) with the number of galaxies within each class included on the top panel (same colour code as lines).
	}
	\label{fig:mass_frac_stats}
\end{figure*}

Following the results presented in the previous sections, let us now argue that the proposed classification based on the ageing diagram is closely connected with the fraction of stellar mass formed during the last \cchange{\tshort~Myr (\massshort) and \tlong~Myr (\masslong)}.

The distribution of \masslong versus \massshort, computed using IllustrisTNG, \pipe+MaNGA, and \pipe+CALIFA star formation histories, is shown on the three panels of Figure~\ref{fig:mass_frac_scatter}, respectively.
We inverted the x-axis order, now displaying high to low fractions, for a more straightforward comparison with the AD.
As in previous figures, each galaxy is coloured according to the AD domain they belong to (Ageing, Undetermined, Quenched and Retired), requiring the same class for both versions of the AD.
If a galaxy does not meet these requirements, it is labelled as unclassified (white points).

We find that, at all panels, the quantities \masslong and \cchange{\massshort} correlate remarkably well with the different AD classes.
On the one hand, Ageing galaxies arrange along a sequence roughly given by a constant SFR over the last \cchange{3~Gyr ($\massshort \simeq 0.067\,\masslong$, illustrated by a black dashed line)}, spanning through a wide range of values of $b$, i.e. current star formation activity compared to the distant past traced by $M_*^{\rm tot}$ (strictly constant $SFR$ over the whole history of the universe, $b=1$ at all times, yields a single point at \cchange{$\massshort \simeq 1.5 \times 10^{-3}$ and $\masslong \simeq 0.22$}).
On the other hand, Retired galaxies correspond to a population that has formed a negligible amount of stars compared to previous epochs, displaying fractions of \cchange{$\masslong\leq 10^{-1}$ and $\massshort \leq 10^{-4}$}.
In contrast, Quenched systems show evidence of having formed a significant fraction of stars during the last 3 Gyr, $\masslong \gtrsim 0.1$.
However, at variance with Ageing galaxies, they have merely formed (if any) less than \cchange{0.1 per cent of the total stellar mass during the last \cchange{\tshort} Myr}.
Finally, the location of Undetermined galaxies in this diagram seems to indicate that they represent an intermediate population between Ageing and Quenched systems.

We conclude thus that the ageing diagram is able to discriminate between different evolutionary scenarios and provide a direct connection between spectral features with physical quantities such as the mass fraction of stars formed on recent timescales.
Once again, we find that IllstrisTNG and \pipe star formation histories yield qualitatively similar results, with the noteworthy exception of \masslong in Quenched and Retired galaxies.
Most simulated galaxies in these categories display null fractions of \massshort (both Q and R) and \masslong (only R), as illustrated by the median values (coloured squares with error bars), while Retired systems in the CALIFA and MaNGA samples have formed at least a few percent ($\sim1-5\%$) of their total stellar mass during the last 3 Gyr according to \pipe \citep[in agreement with independent estimates, e.g.][]{Salvador-Rusinol+20}.

A more through statistical assessment of this discrepancy is shown in Figure~\ref{fig:mass_frac_stats}.
Here, we plot the cumulative distribution of \masslong and \cchange{\massshort} for the three samples, binned in terms of the AD domains.
Regarding IllustrisTNG fractions, we have also included the resulting distribution, considering the formation of an additional stellar particle with $1.4\times10^{6}~\rm M_\odot$ (uniformly distributed within the time after star formation completely stopped; see also Appendix~\ref{appendix:timescales}) in order to investigate the effect of finite numerical resolution.
Although there is good agreement between the simulated and observed distributions regarding Ageing (and possibly Undetermined) galaxies, neither Quenched nor (especially) Retired distributions seem to show statistically compatible results.
However, given the current uncertainties on both the theoretical and observational sides, we simply argue that a more through analysis of the star formation rate in `red-and-dead' galaxies, beyond the reach of the ageing diagram, should be carried out in order to assess this important issue.

\subsection{Comparison with other classification schemes}
\label{sec:sequences}

Our classification scheme indicates the existence of two physically distinct galaxy populations: those that evolve slowly over the entire history of the Universe (Ageing), and those that undergo (fast) quenching, where star formation drops sharply on short ($\le 1$~Gyr) timescales.
If the quenching event took place less than a few hundred Myr ago, these populations are found on the Quenched region of the AD, clearly detached from the Ageing sequence.
From the fraction of galaxies in this region, one may estimate the quenching rate in the local Universe as
	\ccchange{
\begin{equation}
    \eta \sim \frac{f_{\rm Q}}{\tau_{\rm Q}} \approx \frac{[0.03 - 0.11]}{\rm 0.4~Gyr} \sim [0.075 - 0.275]~{\rm Gyr^{-1}},
\end{equation}
where $f_{\rm Q}$ denotes the fraction of galaxies in the Quenched domain (roughly ranging between 3 to 11 per cent according to our results), and $\tau_{\rm Q}$ corresponds to the duration of the Quenched phase, that can be estimated as the difference between the 84 and 16 percentiles of $\tau_{\rm death}$ provided in Table~\ref{tab:tau_death}.
This results are} broadly consistent with recent numerical predictions\footnote{Note that we estimate the quenching rate as $\frac{1}{N} \frac{dN_{\rm Q}}{dt}$, whereas \citet{Wang+21} use $\frac{d(N_{\rm Q}/N)}{dt}$, including the variations $\frac{dN}{dt}$ in the number of galaxies.} \citep{Wang+21}.
Therefore, it seems likely that quenching affects a significant fraction of the galaxy population, if these rates are indeed representative of the cosmic history, as suggested by these authors.

This contrasts, to some extent, with the view that galaxies form a continuous distribution in terms of $M_*-sSFR$  \citep[e.g.][\citetalias{Corcho-Caballero+20}]{Eales+18b}.
In addition to a possible lack of statistics (most galaxies are in the Ageing category in any case), the observed conditional probability distribution of the $sSFR$ is smeared by measurement uncertainties and/or sample selection effects, as well as the intrinsic, physical spread in the residual star formation activity after the quenching event if galaxies do not become completely dead.
The present results highlight that the (specific) star formation rate is, in itself, not sufficient to discriminate Ageing and Quenched galaxies.

\begin{table}
    \centering
    \begin{tabular}{cccc}
    Class & IllustrisTNG & MaNGA & CALIFA\\\hline\hline
    Ageing & 0.50 &  0.54 & 0.75 \\
    Undetermined & 0.02 & 0.01 & 0.00\\
    Quenching & 0.45 & 0.30 & 0.01\\
    Retired & 0.01 & 0.04 & 0.11\\
    Unclassified & 0.02 & 0.11 & 0.13\\\hline
    \end{tabular}
    \caption{Fraction of Green Valley ($0.5 < \gr < 0.7$) galaxies in each AD domain.}
    \label{tab:gv_fractions}
\end{table}

This also applies to the popular classification as `star-forming' or `quenched' based on galaxy colours or $sSFR$ at the time of observation \citep[e.g.][]{Peng+10, Wetzel+13}.
Red galaxies are currently Quiescent in terms of the birthrate parameter and roughly correspond to our Retired population, but they may have reached that point with or without undergoing a quenching event.
The AD suggests that there are two sequences, governed by slow and fast evolution, that extend from blue (high $sSFR$) to red colours (low $sSFR$).
They are well separated at the blue extreme, but their red end-points are indistinguishable.
As it is well known, the interpretation of intermediate sSFRs (e.g. our Transition stage, $0.05 < b < 0.5$) or colours (e.g. the Green Valley, $0.5 < \gr < 0.7$) as `quenching' systems is not well grounded \citep[e.g.][]{Schawinski+14}.
Our results provide further support that these intermediate classes based on optical colour and/or sSFR contain a mixture of Ageing, Quenched, and Retired galaxies, and we argue that our scheme provides a much clearer view of the recent star formation history (see Table~\ref{tab:gv_fractions}).



So far, the best candidates for recently-quenched systems are the so-called Post-Starburst galaxies (PSBGs).
As mentioned in the Introduction, the classical approach to identify these systems is based on the presence of prominent Balmer absorption lines, $\rm EW(H\delta)< 3-5~\AA$, and weak or non-detected \ha emission  \citep[e.g.][]{Dressler&Gunn83, Poggianti+99, Goto+03}, although more sophisticated methods using PCA analysis with spectral and photometric data can also be found \citep[see e.g.][]{Wild+07, Wild+14}.
In the context of the present work, given the strong correlation between $\rm EW(H\delta)$ and \gr or \balbreak \citep[see e.g.][]{Wild+07, Zibetti+17}, PSBGs correspond to the blue extreme of the Quenched sequence in the AD, $\balbreak \lesssim 1.4$.
Our criterion to identify Quenched galaxies is much less restrictive, and therefore it provides a complementary approach to explore the subsequent evolution of PSBGs, as well as to identify quenching events that affect a larger number of galaxies over a broader distribution along the ageing sequence.

\section{Summary and conclusions}
\label{sec:conclusions}
In this work we have studied the connection between the star formation history (SFH) and the location of a galaxy on the \emph{ageing diagram} (AD): the relation between the equivalent width of the \ha line, tracing the presence of young stellar populations, and another probe of star formation on intermediate scales, of the order of $\sim$~Gyr.

We improved upon the methodology originally presented in \citet{Casado+15} and \citet{Corcho-Caballero+21b} by using theoretical predictions from the IllustrisTNG-100 simulation, as well as observational reconstructions of the SFH obtained with \pipe from a sample of MaNGA and CALIFA galaxies.
On the one hand, we use \pipe to derive dust-corrected \gr optical colours, enabling a cleaner separation of star-forming and recently-quenched systems in the AD, as well as a more meaningful comparison between theory and observations.
On the other hand, we also explore the Balmer Break, \balbreak, as an alternative proxy of the star formation activity on $\sim$~Gyr scales, more robust with respect to dust extinction.

We establish a classification in the AD, given by equation~\eqref{eq:ad_fit} with the parameters quoted in Table~\ref{tab:best_params}, that consists of four categories:
\begin{enumerate}
\item 
\textit{Ageing} galaxies ($70-80$ per cent of the population) are systems whose star formation rate changes on scales comparable to the age of the Universe (average $sSFR \sim 10^{-11}-10^{-10}~\rm yr^{-1}$), encompassing from blue (the so-called Main Sequence) to red optical colours.
\item
\textit{Undetermined} systems (a few per cent) represent a transition population caused by recent quenching and/or rejuvenation.
\item
The \textit{Quenched} sequence of galaxies ($5-10\%$) is composed by systems whose star formation activity has abruptly decreased at some point during the last $\sim$~Gyr, with the blue-most end roughly corresponding to classical post-starburst galaxies.
Overall, our Quenched galaxies spend a time $T_{\rm T} \sim 0.5$~Gyr at Transition evolutionary stage (birthrate parameter $0.05 < b < 0.5$), in agreement with previous estimates of the characteristic quenching timescales \citep[e.g.][]{Hahn+17, Wright+19, Tacchella+22}.
In IllustrisTNG, the star formation rate drops to $0$, while \pipe SFHs retain low levels of star formation, with $sSFR \sim 10^{-12}-10^{-11} ~ \rm yr^{-1}$ \citep{Katsianis+21, Corcho-Caballero+21a}.
\item 
\textit{Retired} galaxies ($15-25\%$) distribute over a region of the parameter space where Ageing and Quenched galaxies converge, featuring an optical continuum dominated by old stellar populations.
Regardless of any residual star formation, both \pipe and IllustrisTNG SFHs show that these systems dropped from the Main Sequence ($0.5 < b < 2$) more than 3~Gyr ago.
On average, galaxies in MaNGA and CALIFA samples have spent $\sim 2$ Gyr in the Transition stage over the last 3~Gyr, whereas most simulated galaxies have been completely Quiescent during this period.
\end{enumerate}

As shown in Figure~\ref{fig:mass_frac_scatter}, the observational classification in terms of the AD corresponds to a physical segregation in terms of the stellar mass fractions formed in the last \cchange{\tshort~Myr and 3~Gyr}.
Our main conclusion is that this tool provides a straightforward, robust probe of the star formation history on those time scales, and it cleanly separates galaxies in the Ageing and Quenched sequences.

\section*{Acknowledgements}

\ccchange{We thank the anonymous referee for providing comments that improved the quality of the manuscript.
PCC and YA want to thank Luca Cortese for providing useful comments, as well as financial support provided by grant PID2019-107408GB-C42/AEI/10.13039/501100011033 of the Spanish State Research Agency.
SFS thanks the support by the PAPIIT-DGAPA IG100622 project.}

This research made extensive use of NumPy,\footnote{https://numpy.org/} Matplotlib,\footnote{https://matplotlib.org/} \citep{Hunter:2007} and Astropy,\footnote{http://www.astropy.org} a community-developed core Python package for Astronomy \citep{astropy:2013, astropy:2018}.

This study uses data provided by the Calar Alto Legacy Integral Field Area (CALIFA) survey (http://califa.caha.es/).
This study is based on observations collected at the Centro Astronómico Hispano Alemán (CAHA) at Calar Alto, operated jointly by the Max-Planck-Institut fűr Astronomie and the Instituto de Astrofísica de Andalucía (CSIC).

Funding for the Sloan Digital Sky Survey IV has been provided by the Alfred P. Sloan Foundation, the U.S. Department of Energy Office of Science, and the Participating Institutions. 

SDSS-IV is managed by the Astrophysical Research Consortium for the Participating Institutions of the SDSS Collaboration including the Brazilian Participation Group, the Carnegie Institution for Science, Carnegie Mellon University, Center for Astrophysics | Harvard \& Smithsonian, the Chilean Participation Group, the French Participation Group, Instituto de Astrof\'isica de Canarias, The Johns Hopkins University, Kavli Institute for the Physics and Mathematics of the Universe (IPMU) / University of Tokyo, the Korean Participation Group, Lawrence Berkeley National Laboratory, Leibniz Institut f\"ur Astrophysik Potsdam (AIP),  Max-Planck-Institut f\"ur Astronomie (MPIA Heidelberg), Max-Planck-Institut f\"ur Astrophysik (MPA Garching), Max-Planck-Institut f\"ur Extraterrestrische Physik (MPE), National Astronomical Observatories of China, New Mexico State University, New York University, University of Notre Dame, Observat\'ario Nacional / MCTI, The Ohio State University, Pennsylvania State University, Shanghai Astronomical Observatory, United Kingdom Participation Group, Universidad Nacional Aut\'onoma de M\'exico, University of Arizona, University of Colorado Boulder, University of Oxford, University of Portsmouth, University of Utah, University of Virginia, University of Washington, University of Wisconsin, Vanderbilt University, and Yale University.SDSS is managed by the Astrophysical Research Consortium for the Participating Institutions of the SDSS Collaboration including the Brazilian Participation Group, the Carnegie Institution for Science, Carnegie Mellon University, Center for Astrophysics | Harvard \& Smithsonian (CfA), the Chilean Participation Group, the French Participation Group, Instituto de Astrofísica de Canarias, The Johns Hopkins University, Kavli Institute for the Physics and Mathematics of the Universe (IPMU) / University of Tokyo, the Korean Participation Group, Lawrence Berkeley National Laboratory, Leibniz Institut für Astrophysik Potsdam (AIP), Max-Planck-Institut für Astronomie (MPIA Heidelberg), Max-Planck-Institut für Astrophysik (MPA Garching), Max-Planck-Institut für Extraterrestrische Physik (MPE), National Astronomical Observatories of China, New Mexico State University, New York University, University of Notre Dame, Observatório Nacional / MCTI, The Ohio State University, Pennsylvania State University, Shanghai Astronomical Observatory, United Kingdom Participation Group, Universidad Nacional Autónoma de México, University of Arizona, University of Colorado Boulder, University of Oxford, University of Portsmouth, University of Utah, University of Virginia, University of Washington, University of Wisconsin, Vanderbilt University, and Yale University.

SDSS-IV acknowledges support and resources from the Center for High Performance Computing  at the University of Utah. The SDSS website is www.sdss.org.


\section*{Data Availability}

The data underlying this article are publicly available at https://www.sdss.org/dr17/manga/ for MaNGA and https://califa.caha.es/ for the
CALIFA surveys, respectively.
IllustrisTNG data can be accessed through https://www.tng-project.org/data/.
Additional data generated by the analyses in this
work are available upon request to the corresponding author.



\bibliographystyle{mnras}
\bibliography{bibliography} 


\newpage
\appendix
\section{Timescales across the ageing diagram and resolution effects}
\label{appendix:timescales}
\begin{figure*}
	\includegraphics[width=\linewidth]{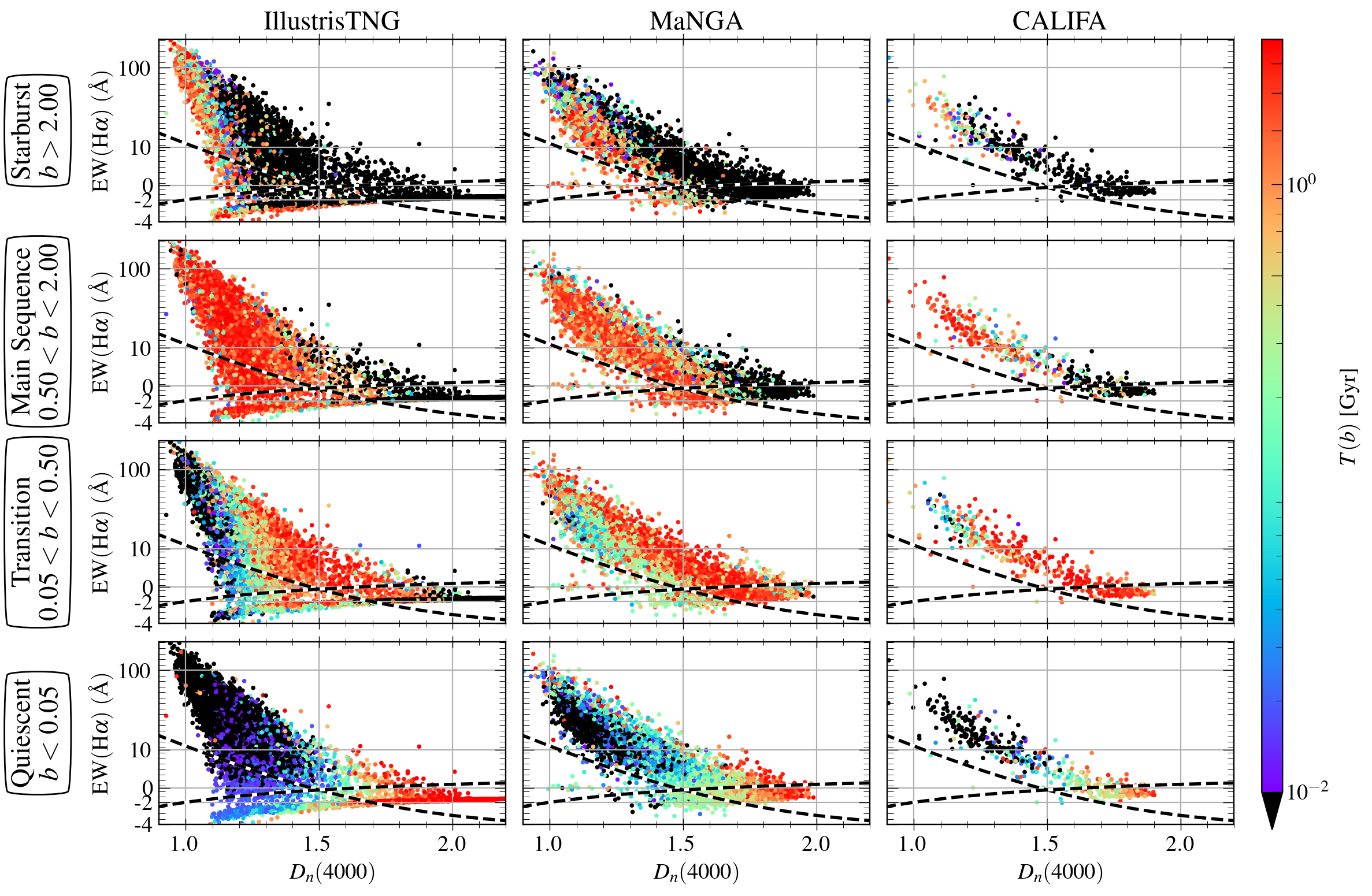}
	\caption{Analogous to Figure~\ref{fig:ad_timescales}, computed in terms of the \balbreak-\ew ageing diagram.}
	\label{fig:ad_timescales_d4000}
\end{figure*}

\begin{figure*}
	\includegraphics[width=\linewidth]{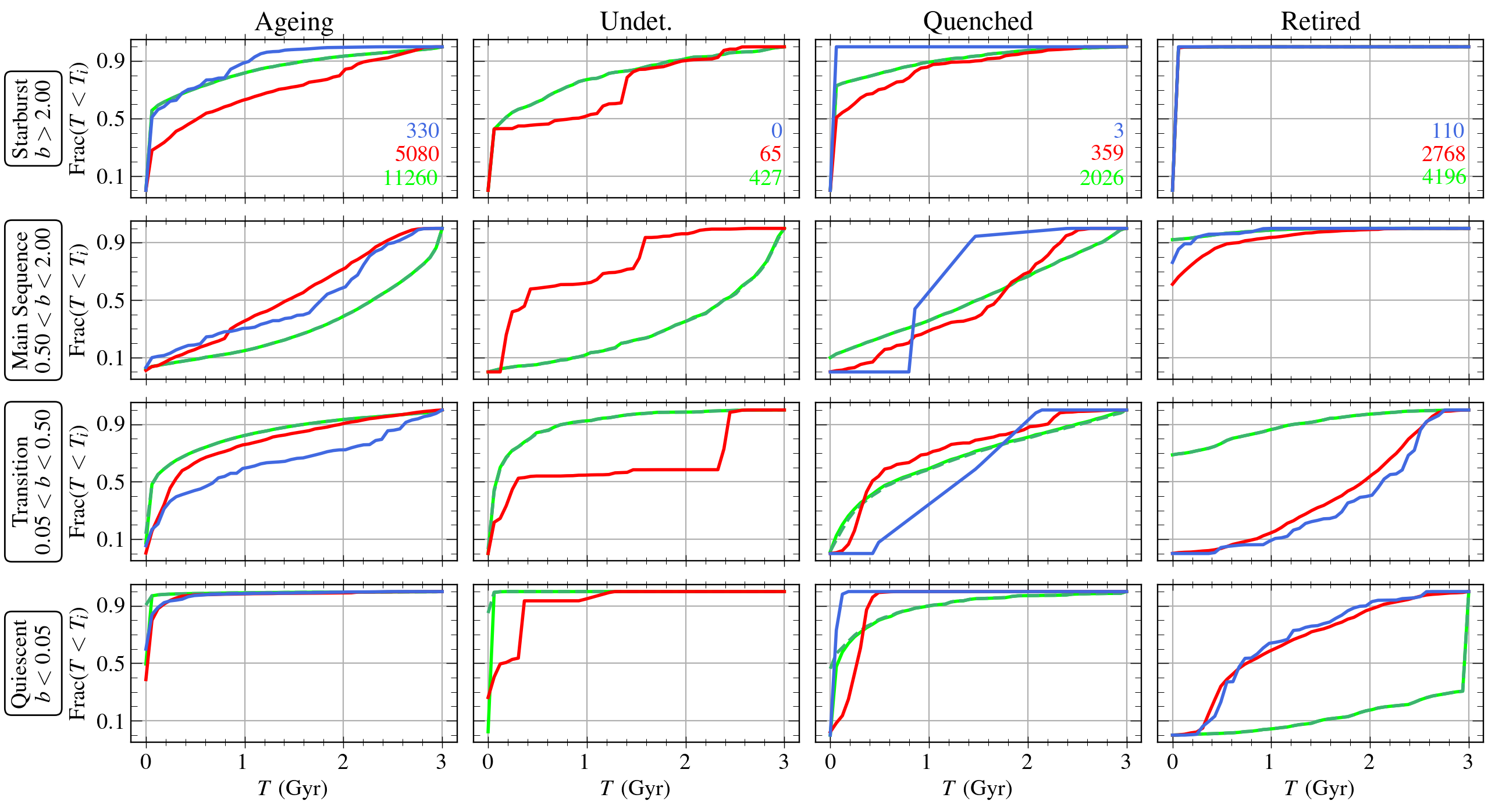}
	\caption{
		Cumulative distribution of time spent, $T$, at each SF activity regime (SB, MS, T, Q), during the last 3 Gyr.
		\ccchange{Results from IllustrisTNG are denoted as light green solid lines, while dark green dashed lines denote the resulting distribution when accounting for resolution effects.}
		The distributions obtained for CALIFA and MaNGA samples are denoted as blue and red solid lines, respectively.
		Each column corresponds to an AD domain (Ageing, Undetermined, Quenched, Retired).
		The number of systems in each domain is shown in the top panel (same colour as lines).}
	\label{fig:T_distrib}
\end{figure*}

In this appendix we include complementary information to support the discussion held in Section~\ref{sec:timescales}.

\subsection{Timescales across \balbreak-\ew}

Figure~\ref{fig:ad_timescales_d4000} displays the distribution of timescales within each SF activity phase in terms of the \balbreak-\ew diagram, analogously to Figure~\ref{fig:ad_timescales}.

\subsection{Statistical distribution of timescales and resolution effects}

In order to provide a better characterisation of the amount of time $T$ that galaxies spend on a given SF regime, we computed the cumulative distribution of $T$ within each phase (SB, MS, T, Q) as a function of AD classification (Ageing, Undetermined, Quenched and Retired).

\cchange{The results are shown in Figure~\ref{fig:T_distrib} for IllustristTNG (green), \pipe+MaNGA (red) and \pipe+CALIFA (blue) star formation histories.}
In addition, we also include for the simulated galaxies the distribution of time (green dashed lines) that would result if they had formed an additional stellar particle ($m_b = 1.4 \times 10^{6}$ M$_\odot$), uniformly distributed after star formation stopped.
In other words, the star formation histories of IllustrisTNG galaxies will have the form:
\begin{eqnarray}
\label{eq:sfh_resolution}
SFR_{\rm TNG}^{+1}(t) = \begin{cases} SFR_{\rm TNG}(t), & \mbox{if } SFR_{\rm TNG}(t) > 0 \\ \frac{m_b}{\tau_{\rm death}}, & \mbox{if } SFR_{\rm TNG}(t>t_0-\tau_{\rm death}) = 0 \end{cases},
\end{eqnarray}
where $SFR_{\rm TNG}(t)$ corresponds to the simulated star formation history (see Sec.~\ref{sec:IllustrisTNG}), $t_0$ denotes the current age of the Universe, and $t_0-\tau_{\rm death}$ corresponds to the cosmic time at which the galaxy formed its last star, i.e. $SFR_{\rm TNG}(t) = 0\ \forall\, t>t_0-\tau_{\rm death}$.

Resolution effects in IllustrisTNG do not seem to play an important role at any SF regime.
In particular, they do not affect the agreement between theory and observations in the Ageing domain nor the discrepancy regarding dead galaxies, that did not form any stars over the last 3~Gyr (about $\sim 70\%$ of the Retired galaxies in IllustrisTNG, and none in the \pipe SFHs).
On shorter time scales, $T$ is still a robust quantity, primarily due to our choice of boundary SF limits (see \S\ref{sec:timescales}).
For example, the time $T(b < 0.05)$ spent at the Quiescent stage is the most sensitive to resolution effects.
Since the birth rate of a dead galaxy given by \eqref{eq:sfh_resolution} is
\begin{equation}
b^{+1}(t > t_0-\tau_{\rm death}) = \frac{m_b}{M_*} \frac{t_0}{\tau_{\rm death}},
\end{equation}
having a birth rate $b \geq 0.05$ at the present time $t_0$ would require $\tau_{\rm death}\leq 0.38$ Gyr for the extreme case of a dwarf galaxy with $M_*=10^9$ M$_\odot$, and even less for higher stellar masses.
Only for such low values of \taudeath would resolution effects bring a dead galaxy into the transition regime $b \geq 0.05$.
The difference between $T_Q$ and $T_Q^{+1}$ will always be less than $380$~Myr, and this correction only affects $\lesssim 30\%$ Quenched and $\lesssim 15\%$ Retired simulated galaxies.
Therefore, we conclude that the time scales $T$ are robust with respect to numerical resolution effects.

\bsp	
\label{lastpage}
\end{document}